\documentclass[aps,prb,twocolumn,amsmath,amssymb,showpacs]{revtex4-2}
\usepackage[utf8]{inputenc}
\usepackage[colorlinks=true, linkcolor=blue, filecolor=blue, citecolor = blue, urlcolor=blue]{hyperref}

\usepackage{amsmath}
\usepackage{amssymb}
\usepackage{amsthm}
\usepackage[utf8]{inputenc}
\usepackage[dvipsnames]{xcolor}

\newcommand{\ket}[1]{| #1 \rangle}
\newcommand{\bra}[1]{\langle #1|}

\newcommand{\tr}{\mbox{Tr}}

\renewcommand{\bf}[1]{\textbf{#1}}
\newcommand{\BZ}{\text{BZ}}

\usepackage[all]{xy}

\usepackage{graphicx}

\usepackage{subcaption}

\begin{document}

\title{Information geometry of quantum critical submanifolds: relevant, marginal and irrelevant operators}

\author{Bruno Mera}
\email{bruno.mera.c5@tohoku.ac.jp}
\affiliation{Advanced Institute for Materials Research (WPI-AIMR), Tohoku University, Sendai 980-8577, Japan}

\author{Nikola Paunkovi\'{c}}
\email{npaunkov@math.tecnico.ulisboa.pt}
\affiliation{Instituto de Telecomunica\c{c}\~oes and Departmento de Matem\'{a}tica, Instituto Superior T\'ecnico, Universidade de Lisboa, Avenida Rovisco Pais 1, 1049-001 Lisboa, Portugal}

\author{Syed Tahir Amin}
\email{tahiramin811@gmail.com}
\affiliation{Instituto de Telecomunica\c{c}\~oes, Avenida Rovisco Pais 1, 1049-001 Lisboa, Portugal;}
\affiliation{Departamento de F\'{\i}sica, Instituto Superior T\'ecnico, Universidade de Lisboa, Av. Rovisco Pais 1, 1049-001 Lisboa, Portugal;}
\affiliation{CeFEMA, Instituto Superior
T\'ecnico, Universidade de Lisboa, Avenida Rovisco Pais 1, 1049-001 Lisboa, Portugal;}
\affiliation{Department of Physics and I3N, University of Aveiro, 3810-193 Aveiro, Portugal;}
\affiliation{Center of Physics of University of Minho and University of Porto, P-4169-007 Oporto, Portugal;}
\affiliation{Department of Physics, Kohsar University Murree, Murree, Pakistan}

\author{V\'{\i}tor R. Vieira}
\email{vitor.rocha.vieira@tecnico.ulisboa.pt}
\affiliation{Departamento de F\'{\i}sica, Instituto Superior
T\'ecnico, Universidade de Lisboa, Avenida Rovisco Pais 1, 1049-001 Lisboa, Portugal}
\affiliation{CeFEMA, Instituto Superior
T\'ecnico, Universidade de Lisboa, Avenida Rovisco Pais 1, 1049-001 Lisboa, Portugal}

\date{\today}

\begin{abstract}
 We analyze the thermodynamical limit of the quantum metric along critical submanifolds of theory space. Building upon various results previously known in the literature, we relate its singular behavior to normal directions, which are naturally associated with relevant operators in the renormalization group sense. We formulate these results in the language of information theory and differential geometry. We exemplify our theory through the paradigmatic examples of the XY and Haldane models, where the normal directions to the critical submanifolds are seen to be precisely those along which the metric has singular behavior, while for the tangent ones it vanishes---these directions lie in the kernel of the metric.
\end{abstract}

\maketitle
\section{Introduction}
\label{sec: introduction}
Classical information geometry is the study of the geometry of statistical manifolds and it finds applications in statistics, information theory, and statistical mechanics~\cite{ama:nag:20,bal:alh:rei:86,mer:mat:car:20}. Recently, information geometry has seen a lot of application in the study of quantum systems and their phase transitions by analyzing different Riemannian metrics over spaces of density operators, a generalization of classical probability distributions~\cite{zan:pau:06, ven:zan:07, zan:ven:07:thermalbures, rez:zan:10, pau:vie:08} (for more details on quantum information geometry and applications, see, for example Refs.~\cite{pet:07, ben:zyc:17,woo:81,bra:cav:94,fuj:nag:95,bro:lan:98,liu:yua:lu:wan:19,sid:jas:kok:20,suz:yan:hay:20}).  Of particular interest are the exotic topological phases of matter that go beyond the standard Landau-Ginzburg classification paradigm and for which there is no well-defined local order parameter, used to infer phase transitions. The quantum metric~\cite{pro:val:80}---a Riemannian metric over the space of pure quantum states based on state distinguishability---was shown to be a quantity able to probe zero-temperature quantum phase transitions among these phases as one varies some parameter in the system~\cite{mer:vla:pau:vie:17,mer:vla:pau:vie:viy:18,ami:mer:vla:pau:vie:18}. In the context of Bloch bands, the quantum metric in momentum space, in the particular setting of flat bands, has received a lot of attention recently, as it gives geometrical contributions to the characterization of a range of different phenomena, such as exotic superconductivity and superfluidity~\cite{peotta2015superfluidity,julku2016geometric,liang2017band,iskin2018quantum}, the stability of fractional Chern insulating phases~\cite{par:roy:son:13, roy:14,cla:lee:tho:qi:dev:15, jackson2015geometric,lee:cla:tho:17,wang2021exact} and light-matter coupling~\cite{topp2021light}. The quantum metric is also central in determining maximally localized Wannier functions~\cite{mar:97, oza:gol:18}, and it can be used as a practical indicator for exotic momentum-space monopoles~\cite{pal:gol:19, sal:gol:pal:20}. The integral of the momentum-space quantum metric is a measure of electron localization, and it can be extracted through spectroscopy measurements~\cite{oza:gol:19}. In two spatial dimensions, one can define an associated complex structure, which is a measure of anisotropy in localization~\cite{mer:20}.

More recently, relations between the quantum metric and the Berry curvature, which determines topological invariants of the system and gives rise to Berry-phase effects, have been established and understood based on the K\"{a}hler structure of the space of quantum states~\cite{oza:mer:21,mer:oza:21}. These relations have come to play an important role in recent studies concerning the so-called ideal Chern bands, which are presumed to be ideal candidates for hosting fractional Chern insulating phases, and the associated band structure engineering and transport~\cite{wang2021exact,mer:tom:21:engineering,wan:liu:21,led:vish:kha:21,tor:peo:bern:21,par:led:kha:etal:21,nor:pal:sturm:21}. More general relations between the quantum metric and topology of quantum states were derived in the context of Dirac Hamiltonians~\cite{mer:zha:gol:21}.

The quantum metric over the space of parameters describing the system, being the ``infinitesimal'' distance between two ground states, allows us to probe phase transitions. As long as the quantum metric is regular, the ground state of the system does not change substantially, keeping the system in the same phase. On the other hand, if the quantum metric becomes singular, it means that the ground state has undergone a dramatic change and thus the system has experienced a phase transition. However, this singular behavior has its subtleties regarding the scaling behavior when performing the thermodynamical infinite volume limit. In the following, we will show how the quantum metric behaves along a submanifold of parameter space composed of critical points---described, in the long wavelength and infinite volume limit, by conformal field theories. We will show that there exist two types of directions of the parameter change: those which take the system away from, and those which move it within the critical submanifold. By performing a scaling analysis, we show that the behaviors of these directions can be interpreted in light of the theory of the renormalization group (RG); namely, the former directions correspond to relevant operators in the quantum theory, while the latter correspond to irrelevant and marginal operators in the quantum theory. While most of these results are known in the literature, they appear to be scattered throughout a number of papers, done by researchers working in different fields of physics who may not necessarily be aware of each other's results. Here, we present a unified formulation of those results by describing them through the language of information theory and differential geometry. We also illustrate our results on the examples of the XY and the Haldane models. Finally, we present conclusions and future lines of research.\\

\section{Quantum metric along a critical submanifold}
\label{sec: Quantum metric along a critical submanifold}
Suppose we are given a many-body Hamiltonian depending smoothly on some parameters, collectively denoted by $x$, living in a smooth manifold $\mathcal{T}$ of dimension $n$. We refer to $\mathcal{T}$ as \emph{theory space}. These parameters will typically correspond to couplings and external fields determining the system's Hamiltonian $H(x)$, with $x\in \mathcal{T}$. We will be interested in the ground state, i.e., the zero-temperature properties of the system. In particular, we consider the pullback of the \emph{Fubini-Study metric} (which is the natural unitarily invariant metric in the space of pure quantum states), also known as the fidelity susceptibility or the information metric, with respect to the family of ground states $\rho(x)=\ket{\psi(x)}\bra{\psi(x)}$ of $H(x)$, with $x\in \mathcal{T}$. Note that there may be parameters for which the ground state is not unique, in which case we will adopt a regularization procedure where we consider the pullback of the \emph{Bures metric}---a Riemannian metric on the space of density operators---under the map $x\mapsto e^{-H(x)/T}/\tr\left(e^{-H(x)/T}\right)$ and eventually take the $T\to 0$ limit. Note that, in order to be able to infer the phase transitions, one must first take the $T\to 0 $ limit before the thermodynamic limit, since the two do not commute, as discussed in Ref.~\cite{ami:mer:vla:pau:vie:18}. Only then is the regularization procedure meaningful giving consistent results and treating non-degenerate and degenerate cases equally. Additionally, the Boltzmann-Gibbs distribution will always treat degenerate ground states equally, because they have the same energy, which is reasonable to assume given no additional information. The Bures metric has two independent contributions coming from the ``classical'' and ``quantum'' parts of the density matrix $\rho=\sum_{j}p_j\ket{\psi_j}\bra{\psi_j}$, corresponding to the variations on the probabilities and the eigenstates of $\rho$, respectively; see Ref.~\cite{zan:ven:07:thermalbures}:
\begin{align}
ds^2=\frac{1}{4}\sum_{j}\frac{dp_j^2}{p_j} +\frac{1}{2}\sum_{j\neq i}|\bra{\psi_j}d\ket{\psi_i}|^2\frac{\left(p_j-p_i\right)^2}{p_i+p_j}.
\end{align}
The first term is the classical Fisher metric corresponding to the statistical model determined by the probability distribution $(p_j)$. The second term, measuring the variations in the eigenstates of $\rho$, has the nice property that for pure states $\rho=\ket{\psi}\bra{\psi}$ it reduces to the Fubini-Study metric
\begin{align}
g_{\text{FS}}=\bra{d\psi}(1-\ket{\psi}\bra{\psi})\ket{d\psi}.
\end{align}
Under the map $x\mapsto\ket{\psi_i(x)}\bra{\psi_i(x)}$, where $\ket{\psi_i(x)}$ is an eigenstate of $H(x)$ (assuming such a state is globally defined, up to a phase), the pullback of the Fubini-Study metric $g_{\text{FS}}$ describes a metric in the space of theories $\mathcal{T}$---the \emph{quantum metric}, which has the nice formula in terms of the remaining eigenstates
\begin{align}
g=\sum_{j\neq i}\bra{d\psi_i(x)}\psi_j(x)\rangle\langle\psi_j(x)\ket{d\psi_i(x)}.
\label{eq: quantum metric}
\end{align}
We will now take $\ket{\psi_i(x)}=\ket{\psi(x)}$ to be the ground state of $H(x)$. It is convenient to interpret the quantum metric $g$ in terms of operators acting on the Hilbert space. At a given point $x\in\mathcal{T}$, the metric assigns a bilinear non-negative pairing $g(u,v)$ of tangent vectors $u,v\in T_{x}\mathcal{T}$. Physically, an element $v\in T_{x}\mathcal{T}$ generates a first order variation in the parameters of the theory at $x\in \mathcal{T}$. As such, it has a corresponding operator, $\mathcal{O}_v$, acting on the Hilbert space, which, using local coordinates $x^i$, $i=1,\dots, n$, where $v=v^i\frac{\partial}{\partial x^i}$, is given by
\begin{align}
\mathcal{O}_{v}=v^{i}\frac{\partial H}{\partial x^i},
\end{align}
where we assume the Einstein summation convention. It is clear that the operator $\mathcal{O}_v$ is associated with perturbation theory in the sense that $H(x+\varepsilon v)= H(x)+\varepsilon \mathcal{O}_v+\text{O}(\varepsilon^2)$. At each parameter point $x$ from our space of theories $\mathcal{T}$, once we fix local coordinates $x^i$, $i=1,\dots, n$, we have a basis for an associated linear space of operators given by $\mathcal{O}_{i}(x)=\mathcal{O}_{\partial/\partial x^{i}}$, $i=1,\dots,n$. The pullback of the Fubini-Study metric under the map $x\mapsto\ket{\psi(x)}\bra{\psi(x)}$ describes the quantum metric in the space of theories $\mathcal{T}$. Note, however, that it may happen that the metric is degenerate, meaning it is not necessarily invertible at every point $x$ where it is defined---simply because the Jacobian of the transformation can be singular. Alternatively, we can view the degeneracy of the metric in light of the tangent-vector-to-operator correspondence $v\mapsto \mathcal{O}_v$. Assuming the eigenstates are nondegenerate in energy, let $\ket{\psi_i(x)}$ be the eigenstates for $H(x)$ with corresponding energies $E_i(x)$. The eigenstates of $H(x +\varepsilon v)$ can be obtained from first-order perturbation theory around $H(x)$ as follows
\begin{align}
\ket{\psi_i(x+\varepsilon v)}=\ket{\psi_i(x)} +\varepsilon \sum_{j\neq i}\frac{\bra{\psi_j(x)}\mathcal{O}_{v}\ket{\psi_i(x)}}{E_{i}(x)-E_j(x)}\ket{\psi_j(x)},
\end{align}
where we assume a parallel transport gauge $\bra{\psi_i(x)}v^k\frac{\partial}{\partial x ^k}\ket{\psi_{i}(x)}=0$. The order $\varepsilon$ term can vanish if $v\in T_{x}\mathcal{T}$ is in the kernel of the assignment $v\mapsto \mathcal{O}_v$. More generally, it can vanish if and only if $\mathcal{O}_v$ preserves the eigenspace $L^{(i)}_x=\text{span}_{\mathbb{C}}\{\ket{\psi_i(x)}\}$, the fiber at $x$ of the $i$th eigenbundle of $H$. Observe that from Eq.~\eqref{eq: quantum metric}, it follows that for any $u\in T_{x}\mathcal{T}$,
\begin{align}
g(v,u)=\frac{1}{2}\sum_{j\neq i}\frac{\bra{\psi_i}\mathcal{O}_{v}\ket{\psi_j}\bra{\psi_{j}}\mathcal{O}_{u}\ket{\psi_i}}{\left(E_i(x)-E_j(x)\right)^2}+\left(v\leftrightarrow u\right),
\end{align}
and we see that $g(v,\cdot)=0$, i.e., $g(v,u)=0$ for all $u\in T_{x}\mathcal{T}$ if and only if  $\mathcal{O}_v$ preserves the eigenspace $L^{(i)}_x$. In short, for the ground state $\rho(x)=\ket{\psi(x)}\bra{\psi(x)}$, the above statement can be expressed as
\begin{align}
[\mathcal{O}_{v},\rho(x)]=0 \iff g(v,\cdot)=0.    
\label{eq: vanishing of the metric}
\end{align}
From a mathematical perspective, we can understand the above statement as follows. There are two points to be considered here. One is that the smooth map $H:\mathcal{T}\ni x\mapsto H(x)$, where $H(x)$ belongs to the (real) vector space of Hermitian operators, has a differential $dH$, and what we call $\mathcal{O}_i$ is the pushforward $dH(\frac{\partial}{\partial x^i})$, which makes sense on a local chart where the local coordinates $x^i$'s are defined. The map $dH$ can be seen as a vector bundle map between the tangent bundle $T\mathcal{T}$ and the pullback under $H$ of the tangent bundle of the (real) vector space of Hermitian operators. The map $dH$ is not necessarily injective, so one must be careful in identifying the tangent space $T_{x}\mathcal{T}$ with the image of $dH$ at a point. 

The second point is that the smooth assignment $P:x\mapsto P(x)=\ket{\psi_i(x)}\bra{\psi_i(x)}$ (provided the $i$th eigenline bundle is well-defined), where we identify the target space as the projectivization of the Hilbert space, also has a differential which is given explicitly by
\begin{align}
dP=QdPP +PdPQ,
\end{align}
where $Q=I-P=\sum_{j\neq i}\ket{\psi_j}\bra{\psi_j}$ is the orthogonal complement projector (here, the $\ket{\psi_j}$'s are locally defined orthonormal eigenvectors of $H$). Above we used the fact that $P^2=P\implies PdPP=0$. Observe that
\begin{align}
QdPP=\sum_{j\neq i}\bra{\psi_j}d\ket{\psi_i}\ket{\psi_{j}}\bra{\psi_i}=\left(PdPQ\right)^\dagger,
\end{align}
and that (under the assumption $E_i\neq E_j$, for $i\neq j$)
\begin{align}
H\ket{\psi_i}=E_i\ket{\psi_i}\implies \frac{\bra{\psi_j}dH\ket{\psi_i}}{E_i-E_j}=\bra{\psi_j}d\ket{\psi_i}.
\end{align}
It follows that
\begin{align}
QdPP=\sum_{j\neq i}\bra{\psi_j}d\ket{\psi_i}\ket{\psi_{j}}\bra{\psi_i}=\sum_{j\neq i}\frac{\bra{\psi_j}dH\ket{\psi_i}}{E_i-E_j}\ket{\psi_j}\bra{\psi_i}.
\end{align}
The differential of $P$, the injectivity of which determines the non-degeneracy of the quantum metric (since the Fubini-Study metric is a Riemannian and hence non-degenerate metric on the projectivization of the Hilbert space), is controlled by the differential of $H$. Observe that for a tangent vector $v$ in $\mathcal{T}$ we have $\left(QdPP\right)(v)=0\iff dP(v) =0$ (because $QdPP$ and $PdPQ$ are orthogonal) and also $dP(v)\iff P^*g_{\text{FS}}(v,\cdot)=0$. Now the only way $\left(QdPP\right)(v)$ vanishes is if $dH(v)=\mathcal{O}_v$ preserves the state $\ket{\psi_i}$. 

We can relate the above discussion to the so-called \emph{symmetric logarithmic derivative equation} as follows. The symmetric logarithmic derivative of a family of density operators is determined by an operator valued $1$-form $G$ solving the symmetric logarithmic derivative equation
\begin{align}
\label{eq:sld}
d\rho= G\rho +\rho G.
\end{align}
In the particular case of $\rho=P=\ket{\psi_i}\bra{\psi_i}$, we can see that 
\begin{align}
G=QdPP+PdPQ
\end{align}
solves Eq.~\eqref{eq:sld} and, hence, the vanishing of $\left(QdPP\right)(v)$, for a tangent vector $v$, is equivalent to the vanishing of $G(v)$, known as the symmetric logarithmic derivative operator of $\rho=P$ with respect to $v$.

\begin{figure}[h!]
    \centering
    \includegraphics[scale=0.8]{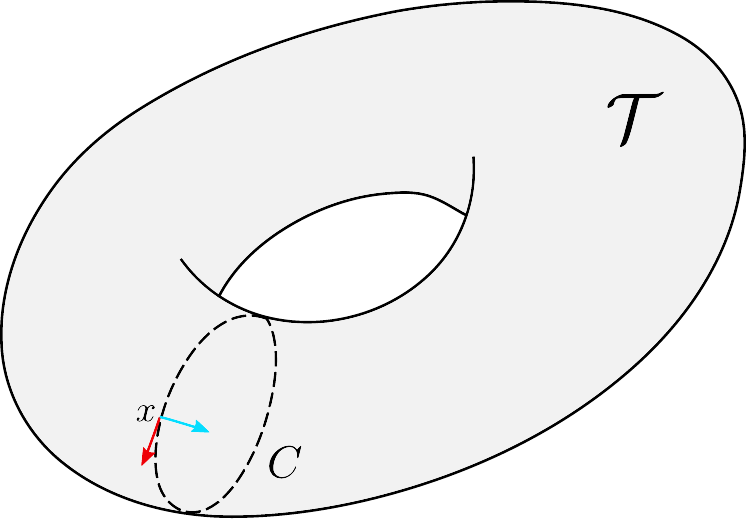}
    \caption{An illustration of the space of theories $\mathcal{T}$ (here, a compact surface of genus $1$), together with a quantum critical submanifold $C\subset \mathcal{T}$ (here, a circle). The red arrow is meant to illustrate a tangent vector to $C$ at $x\in C$, which maps to a marginal or irrelevant operator. The blue arrow illustrates a normal vector to $C$ at $x\in C$, mapping to a relevant operator.}
    \label{fig: illustration}
\end{figure}

Suppose now we have a set $C\subset \mathcal{T}$ of \emph{quantum critical points} where the theory is gapless and our geometric quantity of interest $g$ is presumably singular in the thermodynamical limit. Note that the notion of singularity here is different from the familiar notion of singularity of a smooth vector field in the context of differential geometry. Here the singularity is associated with the quantum metric. More precisely, we say that the quantum metric has a singularity at a point $x\in\mathcal{T}$ if, by taking local coordinates centered at $x$ and hence obtaining a matrix representation of it (by considering the natural coordinate tangent vectors), some of the associated matrix elements are not defined at that point. Note that this notion is really coordinate independent: If this happens in a coordinate system, it will happen in any coordinate system. The thermodynamical limit assumes, in particular, that the volume of the system $L^d$, where $L$ is the linear length of the system and $d$ is the spatial dimension, goes to infinity. For considerations involving the thermodynamical limit, it will be convenient to consider the rescaled quantum metric $g/L^d$. We also make the assumption that the quantum critical theories, as described by points $x\in C\subset \mathcal{T}$, are conformally invariant so that they are described, in the continuum limit and at large distances, by a conformal field theory ($\text{CFT}_d$). For simplicity, we assume that $C$ is a submanifold of $\mathcal{T}$, and we refer to it as a \emph{critical submanifold}. Note that in the model examples below there will be regions of theory space which are critical in the same sense, i.e., the Hamiltonian is gapless in those regions, however these regions will fail to be submanifolds. We will refer to these regions as \emph{critical regions}. Of particular interest to us will be the tangent bundle of $C$ in $\mathcal{T}$, i.e., the vector bundle over $C$ that at each point $x\in C$ associates the tangent space $T_{x}C\subset T_{x}\mathcal{T}$. The point that we want to make, which will be clear from the discussion below, is that the rescaled metric $g/L^d$ is actually, in the thermodynamical limit, only divergent along directions which are complementary to $TC$, which we refer to as directions of \emph{increased ground-state distinguishability} and are associated with \emph{relevant operators} in the framework of the RG. To be precise, the thermodynamical limit of $g/L^d$, which we denote by $g_{*}$, vanishes exactly along tangent vectors to $C$, i.e., $g_*(v,\cdot)=0$ for any vector $v\in T_{x}C\subset T_{x}\mathcal{T}$. As a consequence, the directions where $g_*$ is nonvanishing are normal to $TC$---i.e., each can be seen, effectively, as orthogonal to $TC$ with respect to some Riemannian metric in $\mathcal{T}$. The rigorous statement is that, for each $x\in C$, there is an \emph{exact sequence of vector spaces}
\begin{align}
0\longrightarrow T_{x}C \longrightarrow T_{x} \mathcal{T}\longrightarrow N_{x}C\longrightarrow 0,
\end{align}
defining the vector space of normal directions at $x$, $N_{x}C=T_{x}\mathcal{T}/T_{x}C$, giving rise, globally, to the normal bundle $N C=T\mathcal{T}|_{C}/TC\to C$, see Fig.~\ref{fig: illustration} (by an exact sequence of vectors spaces, we mean a sequence of vector spaces and linear maps where the kernel of the next is equal to the image of the current). Observe that for $x\in C$ and any $u\in T_{x}\mathcal{T}$, if $g_*(u,u)\neq 0$ we have that $g_*(u+v,u)=g_*(u,u)$ for any $v\in T_{x}C$, since $v$ is annihilated by $g_{*}$. The vector $u$ defines a nontrivial element $[u]$ of $N_{x}C=T_{x}\mathcal{T}/T_x C$ and $g_*(u,u)$ only depends on $[u]$. A choice of a complement of $T_xC$ gives a particular unique representation of $[u]$ as an element of $T_x\mathcal{T}$. In practice, one chooses an auxiliary Riemannian metric $h$ defined over $\mathcal{T}$ (actually, it is enough to have it defined over $T\mathcal{T}|_{C}$) and identifies $N_{x}C\cong \left( T_{x}C\right)^{\perp_h}$ using the notion of an orthogonal complement provided by $h$. Importantly, the isomorphism class of this vector bundle is independent of this choice. Finally, we remark that, in light of Eq.~\eqref{eq: vanishing of the metric}, we see that the operators associated with $T_xC$ commute with the ground state at $x$ modulo subextensive terms, i.e., terms that scale as $L^{r}$, with $r<d$.\\

For the convenience of the reader, before moving on to the relation with RG, we list our assumptions\\
\begin{itemize}
    \item [(i)] The theory space or parameter space $\mathcal{T}$ is assumed to be a smooth ($C^{\infty}$) manifold.
    \item [(ii)] A critical submanifold $C$ is a submanifold $C\subset \mathcal{T}$ where $H(x)$ is, in the thermodynamical limit, gapless for every $x\in C$. If $C$ is not a submanifold (as happens in the examples considered) but $H(x)$ is, in the thermodynamical limit, gapless for every $x\in C$, we refer to it as a critical region.
    \item [(iii)] Points in $C$ are assumed to be described by conformal field theories in the continuum limit, at low energies.
\end{itemize}

\section{Relation to RG}
\label{sec: relation to RG}
We consider a critical point $x_0\in C\subset\mathcal{T}$ and assume that the $\mathcal{O}_i$'s are local operators, meaning that they can be written as
\begin{align}
\mathcal{O}_i=\sum_{\bf{r}}\mathcal{O}_i(\bf{r}),
\end{align}
for $\mathcal{O}_i(\bf{r})$ having support in some bounded neighborhood of $\bf{r}$. The sum over lattice sites $\bf{r}$ makes sense before taking the continuum limit. It is convenient to introduce the imaginary-time-evolved operators, namely,
\begin{align}
\mathcal{O}_i(\tau,\bf{r})=e^{\tau H(x_0)} \mathcal{O}_i(\bf{r})e^{-\tau H(x_0)},
\end{align}
where $\tau$ denotes the imaginary time. At the critical point $x_0\in C$, we may assume that the $\mathcal{O}_i$'s have well-defined scaling dimensions $\Delta_i$ such that 
\begin{align}
\mathcal{O}_i(\tau,\bf{r})\longrightarrow \mathcal{O}'_i(\tau,\bf{r})=\mathcal{O}_i(\zeta^{z}\tau,\zeta\bf{r})\zeta^{\Delta_i},
\end{align}
under a transformation $\bf{r}\to \zeta\bf{r} $ and $\tau\to \zeta^{z}\tau $, where $z$ is the dynamical critical exponent. If this is not the case, assuming $\dim \mathcal{T}$ is big enough, we may use a linear coordinate change, so that the above equation holds. These scaling dimensions arise from the linearization of the renormalization group flow near the critical point $x_0\in C$; see Ref.~\cite{ton:17} for details.

Due to the expression
\begin{align}
\label{eq: 2pointcorr and metric}
g_{ij}(x)=&\int_{0}^{\infty}d\tau \tau e^{-\varepsilon \tau}\nonumber\\
&\times \left[\frac{1}{2}\langle \{ \mathcal{O}_i(\tau) ,\mathcal{O}_j(0) \}\rangle -\langle \mathcal{O}_i(\tau) \rangle\langle \mathcal{O}_j(0)\rangle\rangle\right],      
\end{align}
with $\varepsilon\to 0^+$ and $\{\cdot,\cdot\}$ being the anticommutator, one discovers that $g_{ij}(x_0)$ has the finite size scaling law originally found by Campos Venuti, Zanardi, and co-workers~\cite{ven:zan:07,rez:zan:10} and later found in the work of Miyaji \emph{et al.}~\cite{miy:15}, in the context of gauge-gravity duality,
\begin{align}
g_{ij}(x_0)\sim L^{2d+2z -\Delta_{i}-\Delta_{j}},
\end{align}
where $L$ is the linear size of the system and $d$ is its spatial dimension. In particular, we have the following classification of operators according to how their couplings are renormalized under an RG transformation, as we flow onto the infrared energy scales,
\begin{itemize}
    \item [(i)] $\mathcal{O}_i$ is \emph{relevant} if $\Delta_i<d+z$,
    \item [(ii)] $\mathcal{O}_i$ is \emph{marginal} if $\Delta_i=d+z$,
    \item [(iii)] $\mathcal{O}_i$ is \emph{irrelevant} if $\Delta_i>d+z$.
\end{itemize}
We then see that for tangent vectors whose associated operators are \emph{relevant} operators, the metric should blow up in the $L\to\infty$ thermodynamical limit. For tangent vectors associated with \emph{irrelevant} operators the metric should vanish in the thermodynamical limit. Finally, for tangent vectors associated with \emph{marginal} operators, the metric should scale as $L^0$, meaning it should be finite, with possible logarithmic corrections. Accordingly, if we are in a point $x$ taken from  a small neighborhood of $x_0\in C$, the relevant operators move us outside $C$, the irrelevant operators move us back to $x_0$, while the marginal operators move us within $C$, since the corresponding theory is also \emph{scale invariant}. Thus, we expect that the tangent space $T_{x_0}C$ can be identified with the subspace of all irrelevant and marginal operators in the image of $T_{x_0}\mathcal{T}\ni v\mapsto \mathcal{O}_v$. The marginal operators in the image of $T_{x_0}\mathcal{T}\ni v\mapsto \mathcal{O}_v$ then form a complementary subspace $N_{x_0}C$, such that $T_{x_0}\mathcal{T}=T_{x_0}C\oplus N_{x_0}C$. 

Consider the rescaled metric $g_{ij}(x_0)/L^{d}$. The scaling behavior of its diagonal components is given by
\begin{align}
\frac{g_{ii}}{L^d}\sim    \frac{1}{L^d}L^{2z+2d-2\Delta_i},
\end{align}
and hence it vanishes for $2z + d - 2\Delta_{i}<0$ and is finite or blows up for $2z + d - 2\Delta_{i}\geq 0$ in the thermodynamic limit. In particular, for irrelevant and marginal operators, it goes to zero. Surprisingly, it also vanishes for those relevant operators that are not ``sufficiently relevant'', namely those that satisfy
\begin{align}
z+d>\Delta_{i}> z+\frac{d}{2}.   
\end{align}
Its thermodynamical limit $g_{*}$ vanishes along $T_{x_0}C$, so we can identify $T_{x_0}C$ as a subspace of $T_{x_0}\mathcal{T}$ where the rescaled metric has a nontrivial kernel. In the following, we consider two particular examples for which the tangent bundle to $C$ in $\mathcal{T}$ is exactly the kernel of $g_*$---the XY and the Haldane models.
\\
\section{XY model}
\label{sec: XY model}
In the XY anisotropic spin-half chain with $N$ sites on a circle in the presence of an external magnetic field, we have the family of Hamiltonians
\begin{align}
H(\gamma,\lambda) = - \sum_{j=0}^{N-1} \left(\frac{1+\gamma}{2}\sigma^{x}_{j}\sigma_{j+1}^x +\frac{1-\gamma}{2}\sigma_{j}^y\sigma_{j+1}^y +\frac{\lambda}{2}\sigma_j^z\right),
\end{align}
parametrized by $(\gamma,\lambda)\in \mathbb{R}^2$, where $\gamma$ is the anisotropy and $\lambda$ the magnetic field. This model is usually solved by using a Jordan-Wigner transformation,
\begin{align}
c_j=e^{i\pi\sum_{k=0}^{j-1}\sigma^{+}_j\sigma^{-}_j}\sigma_{j}^{-} \text{ and } c_j^{\dagger}=\sigma_{j}^{+}e^{-i\pi \sum_{k=0}^{j-1}\sigma^{+}_j\sigma^{-}_j},
\end{align}
which takes spin variables $\sigma_j$'s to fermionic ones $c_j$'s, to get
\begin{align}
&H(\gamma,\lambda)\nonumber \\
&= -\sum_{j=0}^{N-1}\left[ \left(c_{j+1}^\dagger c_{j} +c_{j}^\dagger c_{j+1}\right)+\gamma\left(c_{j+1}c_{j}+c_{j}^\dagger c_{j+1}^\dagger\right)\right] \nonumber \\
&-2\lambda\sum_{j=0}^{N-1}\left(c_j^\dagger c_j -\frac{1}{2}\right),
\end{align}
where the fermions satisfy twisted boundary conditions $c_{N}^\dagger=c_0^\dagger (-1)^{\sum_{j}c_j^\dagger c_j}$ according to the parity operator.

One then considers the system as a whole with fixed periodic or antiperiodic boundary conditions for the fermions. The translation invariance of the Hamiltonian motivates us to Fourier-expand $c_{j}^\dagger=\frac{1}{\sqrt{N}}\sum_{k} e^{-ik j}c_{k}^\dagger$, where, due to the boundary conditions, we have $e^{ik N}=\pm 1$, which gives the two sets of allowed momenta. For the periodic boundary conditions the allowed momenta have the form
\begin{align}
k=\frac{2\pi}{N}m,\ m\in \{0,\dots, N-1\},
\end{align}
while for the antiperiodic boundary conditions we have
\begin{align}
k=\frac{2\pi}{N}m +\frac{\pi}{N},\ m\in\{0,\dots,N-1\}.
\end{align}
In either case, the corresponding two Hamiltonians assume the same form
\begin{align}
H(\gamma,\lambda) = &\sum_{k}\Big[\left(-\lambda-\cos(k)\right)\left(c_{k}^\dagger c_k-c_{-k}c_{-k}^\dagger\right) \nonumber \\
&+i\gamma \sin(k) c_{-k}c_{k} -i\gamma \sin(k) c_{k}^\dagger c_{-k}^{\dagger}\Big],
\end{align}
where we dropped the overall constant $-\sum_{k}\cos(k)$. It is convenient to introduce the Nambu spinor
\begin{align}
\psi_{k}^\dagger=\left( c_{k}^\dagger \ c_{-k}\right),
\end{align}
in terms of which we have
\begin{align}
H(\gamma,\lambda) = \frac{1}{2}\sum_{k}\psi_{k}^\dagger H(k)\psi_k,
\end{align}
with $H(k)=\vec{d}(k)\cdot \vec{\sigma}$ and $\vec{d}(k)=\left(0,2\gamma \sin(k), -2\left(\lambda+\cos(k)\right)\right)$. One can then diagonalize each Hamiltonian through a Bogoliubov-Valatin transformation to obtain the two spectra and corresponding eigenstates. By fixing periodic boundary conditions for the fermions, only the states with even parity are true eigenstates of the original Hamiltonian. Similarly, fixing anti-periodic boundary conditions, only odd parity states are eigenstates of the original Hamiltonian. To find the true ground state of the system one needs to consider the lowest energy state according to this prescription, the parity of which may depend on the considered system size. In particular, it is known (see Ref.~\cite{ara:mat:85}) that there are phases with two degenerate ground states for $|\lambda| < 1$ and $\gamma \neq 0$, and even phases with infinitely many ground states in the thermodynamical limit (where $\lambda = 0$ and $\gamma = \pm 1$).

Instead of considering this procedure, because solving the XY model is not the aim of this paper, for simplicity, we will fix periodic boundary conditions for the fermions and take that as our model. The ground state is then easy to understand, and one can perform the standard analysis in the thermodynamic limit. The critical region then consists of three critical submanifolds given, respectively, by the segment defined by $\gamma = 0$ and $|\lambda| \leq 1$, and the two lines given by $|\lambda| = 1$ (see for example Ref.~\cite{zan:pau:06}). The segment and the lines intersect transversally. The whole set, i.e., the union of the critical submanifolds, is not a submanifold itself, as the intersection points do not have neighborhoods homeomorphic to $\mathbb{R}$.

For this model, we see that $g_{*}$ restricted to the critical lines vanishes precisely on tangent vectors to the critical lines, except on the intersections $(\lambda,\gamma)=(\pm 1,0)$, where it blows up in all directions. On the complementary subspaces to the tangent spaces it always blows up. In this case there are no operators which are not sufficiently relevant. We remark that at the two critical points $(\gamma,\lambda)=(0,\pm 1)$, corresponding to the intersection between the segment and the two lines, the metric blows up along any direction and, hence, they behave as if they were two isolated critical points (because the normal spaces in this case coincide with the whole of $T_{(0,\pm 1)}\mathcal{T}$). Note, however, that they are not isolated critical points in the strict sense, as any neighborhood of them finds other critical points. We refer the reader to the Appendix for details.\\
\section{Haldane model}
\label{sec: Haldane model}
We consider the Haldane model Hamiltonian~\cite{hal:88}
\begin{equation}
\mathcal{H} \! = \! ~t\!\! \sum_{\langle l,m\rangle}c_{l}^{\dagger}c_{m} + t' \!\!\! \sum_{\langle\langle l,m\rangle\rangle} \!\! e^{-i\nu _{lm}\phi}c_{l}^{\dagger}c_{m} + M\sum_{l}\epsilon_{l} c_{l}^{\dagger}c_{l}, 
\label{eq:2DTS_longrange}
\end{equation}
where $\epsilon_{l}=1$ on site A and $-1$ on site B; $\nu _{lm}=\pm 1$ depending on the direction of next nearest neighbor hopping. This model supports quantized conductance without applying an external magnetic field.  We also fix $t=1$ and $t'=1/3$.  In this case, $(\phi,M)$ are the coordinates describing the parameter manifold in which different topological phases exist. We remark that, in this case, the manifold has the topology of an infinite cylinder $S^1\times \mathbb{R}$, since we have to identify $\phi$ and $\phi+2\pi$. The critical region for this model is well-known and in the $(\phi,M)$ plane they have the shape of a \emph{figure eight}; see~\cite{hal:88}. Once we identify the points $\phi\sim \phi+2\pi$, the resulting critical region consists of two circles which intersect at two points described by $(M,\phi)=(0,0)$ and $(\pi,0)$, both intersecting transversally. To calculate the fidelity susceptibility, one considers the model on a two-dimensional (2D) lattice with $L^2$ sites and takes periodic boundary conditions. Then, one can write the Hamiltonian, in momentum space, which is described by a $2\times 2$ Bloch Hamiltonian of the form $H(\bf{k})=\vec{d}(\bf{k})\cdot \vec{\sigma}$, where $\vec{d}(\bf{k})$ is a three dimensional vector, $\vec{\sigma}$ is the vector of Pauli matrices, and $\bf{k}$ is the momentum in the first Brillouin zone $\BZ^2$, see Ref.~\cite{hal:88} [we are omitting a term proportional to the identity matrix which is irrelevant for the present discussion]. Explicitly, the vector $\bf{d}(\bf{k})=\left(d_1(\bf{k}),d_2(\bf{k}),d_3(\bf{k})\right)$ is given by
\begin{align}
d_1(\bf{k}) =& 1+ \cos(\bf{k}\cdot \bf{a}_1) +\cos(\bf{k}\cdot\bf{a}_2),\nonumber \\
d_2(\bf{k}) =&\sin(\bf{k}\cdot \bf{a}_1) +\sin(\bf{k}\cdot\bf{a}_2), \nonumber \\
d_3(\bf{k}) =&M +\frac{2}{3}\sin(\phi)\Big[\sin(\bf{k}\cdot\bf{a}_1)-\sin(\bf{k}\cdot\bf{a}_2)\nonumber\\
&-\sin\big(\bf{k}\cdot(\bf{a}_1-\bf{a}_2)\big)\Big], 
\end{align}
with $\bf{a}_1=(a/2)(3,\sqrt{3})$ and $\bf{a}_2=(a/2)(3,-\sqrt{3})$, where $a$ is the lattice constant that we take to be equal to $1$.
The quantum metric over the parameter space, is then given~by
\begin{widetext}
\begin{align}
g(M,\phi) = \frac{1}{4}\sum_{\bf{k}} \left(\frac{\partial \vec{n}}{\partial M}\cdot \frac{\partial \vec{n}}{\partial M}dM^2+2\frac{\partial \vec{n}}{\partial M}\cdot \frac{\partial \vec{n}}{\partial \phi}d\phi dM +\frac{\partial \vec{n}}{\partial \phi}\cdot \frac{\partial \vec{n}}{\partial \phi}d\phi^2\right),     
\end{align}
\end{widetext}
where $\vec{n}(\bf{k})= \vec{d}(\bf{k})/|\vec{d}(\bf{k})|$, and the sum is restricted to the allowed momenta for the corresponding finite-size system which yields a discrete torus inside the Brillouin zone---more explicitly, these momenta are represented in $\mathbb{R}^2$ by $\bf{k}=\sum_{i=1}^2k_i\bf{e}_i$, where $k_i\in\{\frac{2\pi j}{L}:j=0,\dots,L-1\}$, $i=1,2$, and $\{\bf{e}_i\}_{i=1}^{2}$ is a basis for the reciprocal lattice. The fidelity susceptibility is then $\chi= g/L^2$. We remark that the expression for $\chi$ defines a Riemann sum for an integral and, once we take $N\to\infty$, we obtain the formula
\begin{widetext}
\begin{align}
\chi(M,\phi)= \frac{1}{4}\int_{\BZ^2}\frac{d^2\bf{k}}{\left(2\pi\right)^2} \left(\frac{\partial \vec{n}}{\partial M}\cdot \frac{\partial \vec{n}}{\partial M}dM^2+2\frac{\partial \vec{n}}{\partial M}\cdot \frac{\partial \vec{n}}{\partial \phi}d\phi dM +\frac{\partial \vec{n}}{\partial \phi}\cdot \frac{\partial \vec{n}}{\partial \phi}d\phi^2\right),
\end{align}
\end{widetext}
which can safely be used away from the critical points. For numerical purposes, one can use the procedure outlined in  Sec.~IV of Ref.~\cite{ami:mer:vla:pau:vie:18} to compute $g$ and $\chi$ for a finite system and a very small, but nonvanishing, temperature.

\begin{figure}[h]
    \centering
    \begin{subfigure}{0.2\textwidth}
    \centering
    \includegraphics[width=\textwidth]{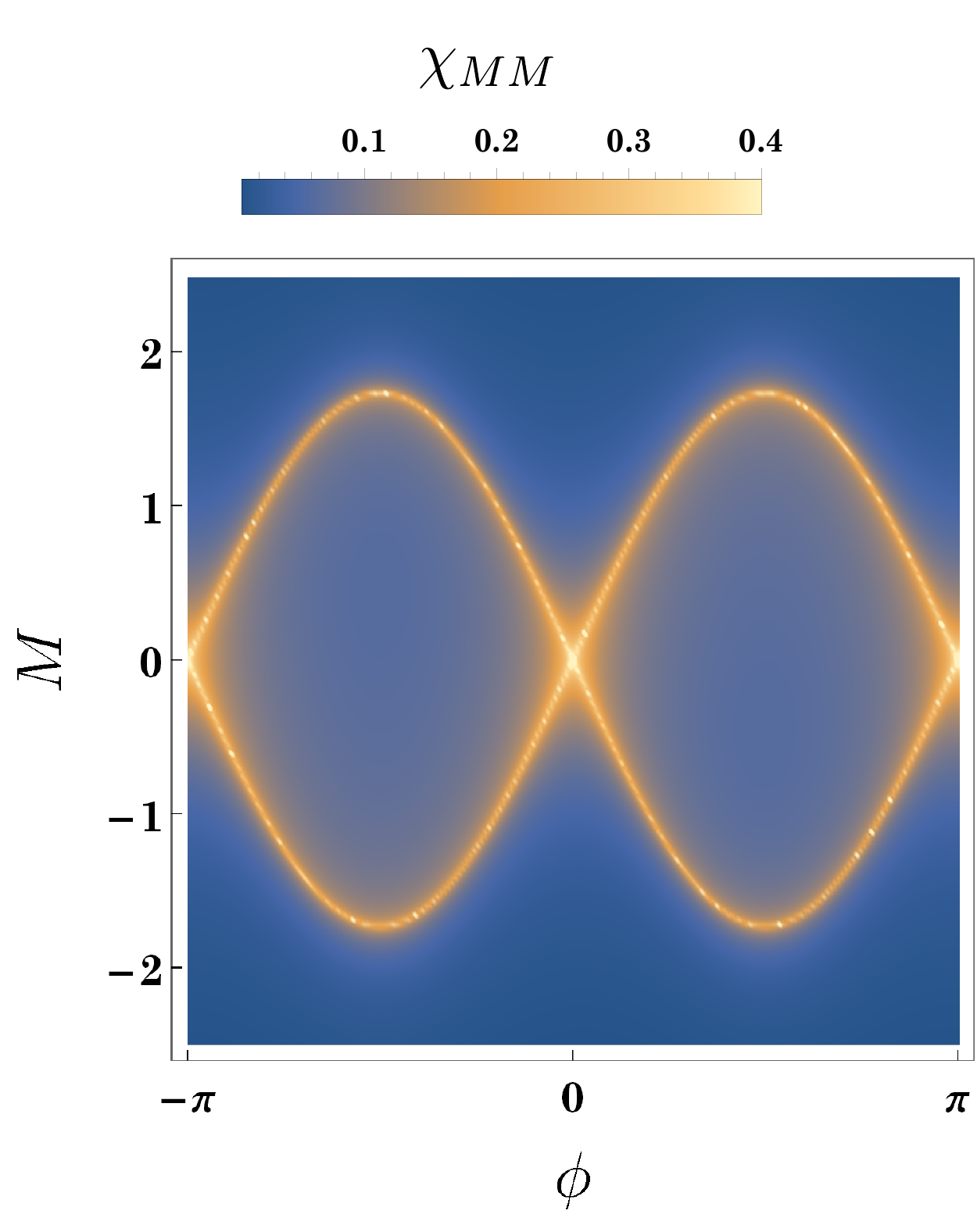}
    \end{subfigure}
    \begin{subfigure}{0.2\textwidth}
    \centering
    \includegraphics[width=\textwidth]{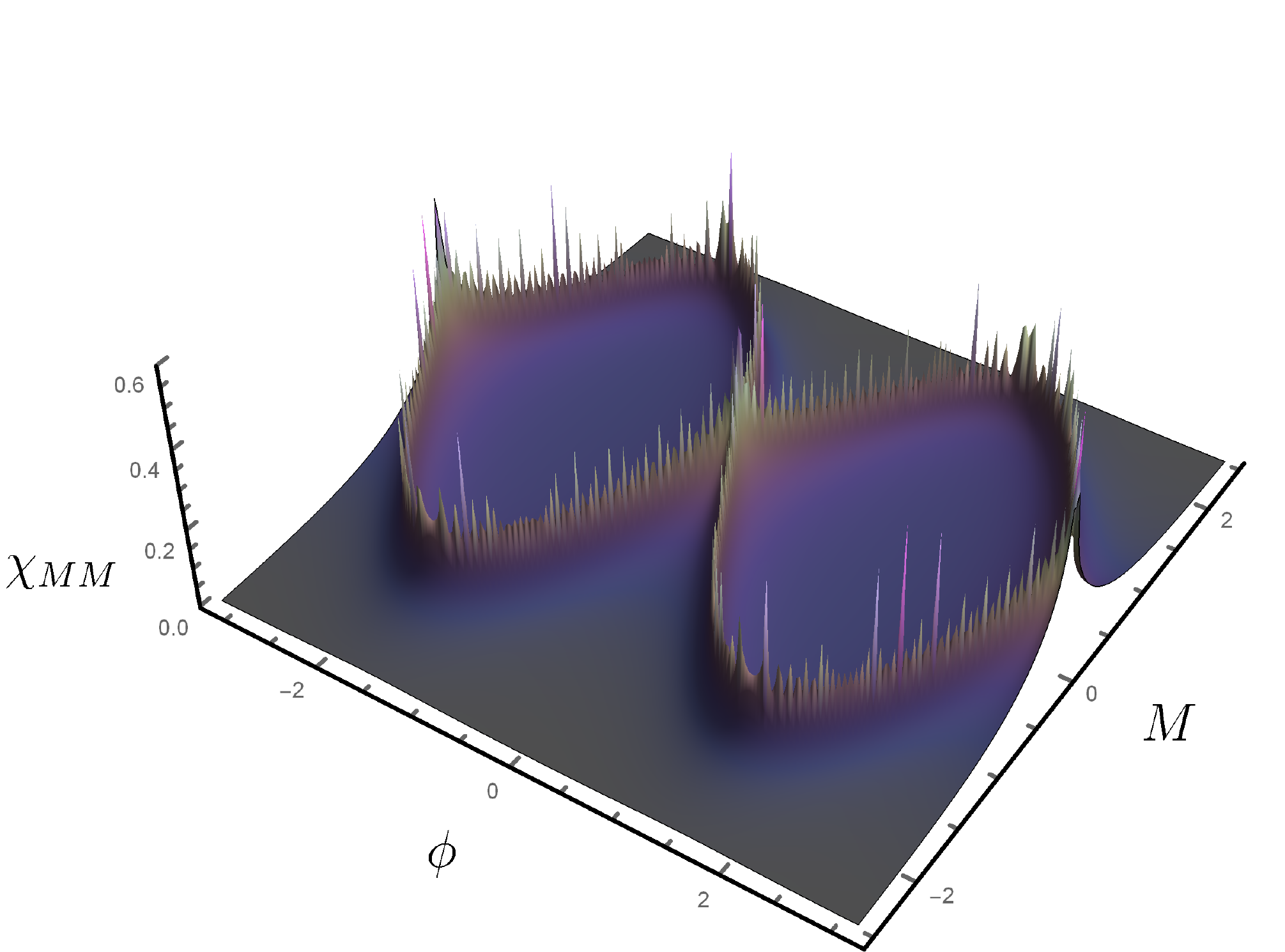}
    \end{subfigure}
    \begin{subfigure}{0.2\textwidth}
    \centering
    \includegraphics[width=\textwidth]{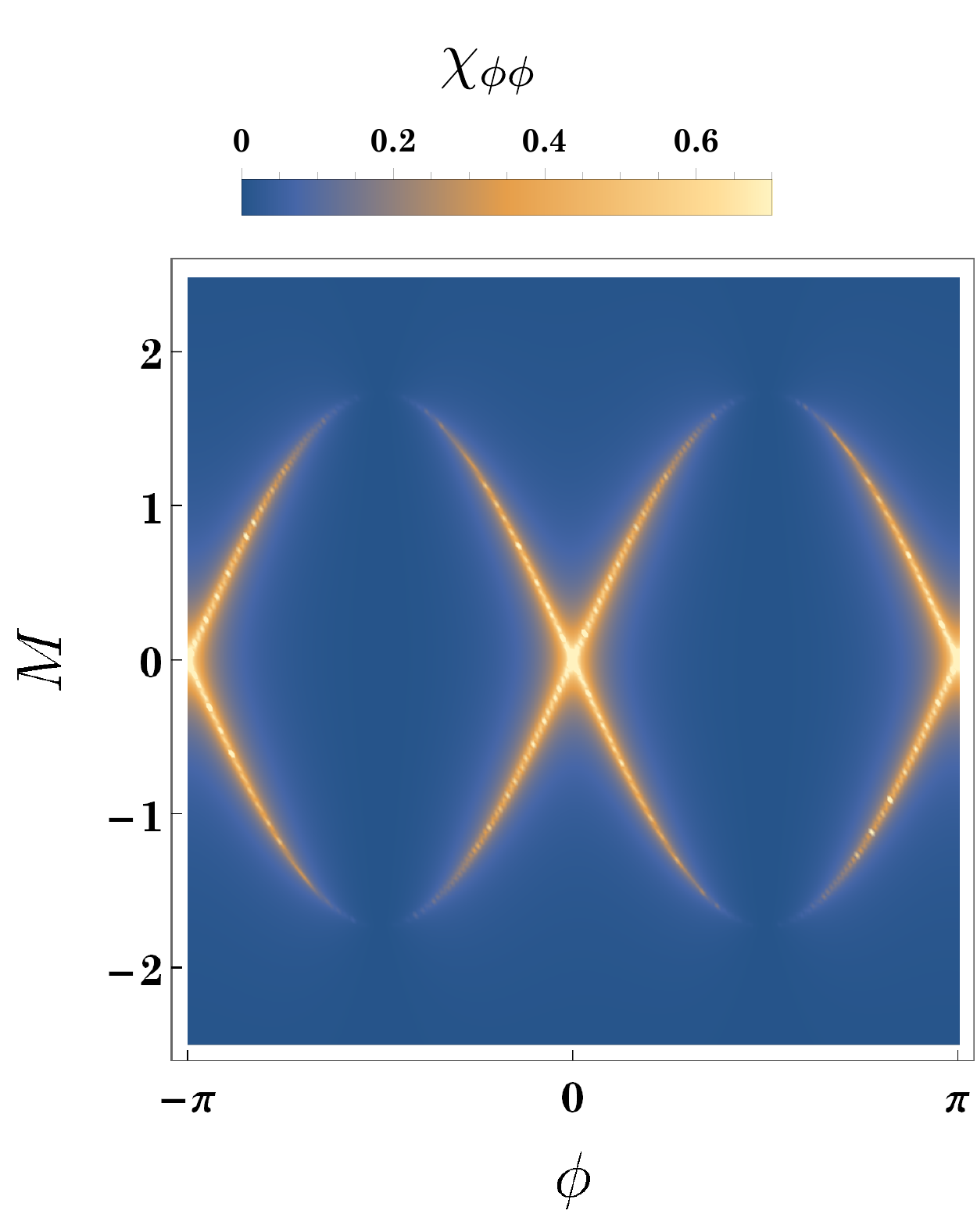}
    \end{subfigure}
   \begin{subfigure}{0.2\textwidth}
   \centering
   \includegraphics[width=\textwidth]{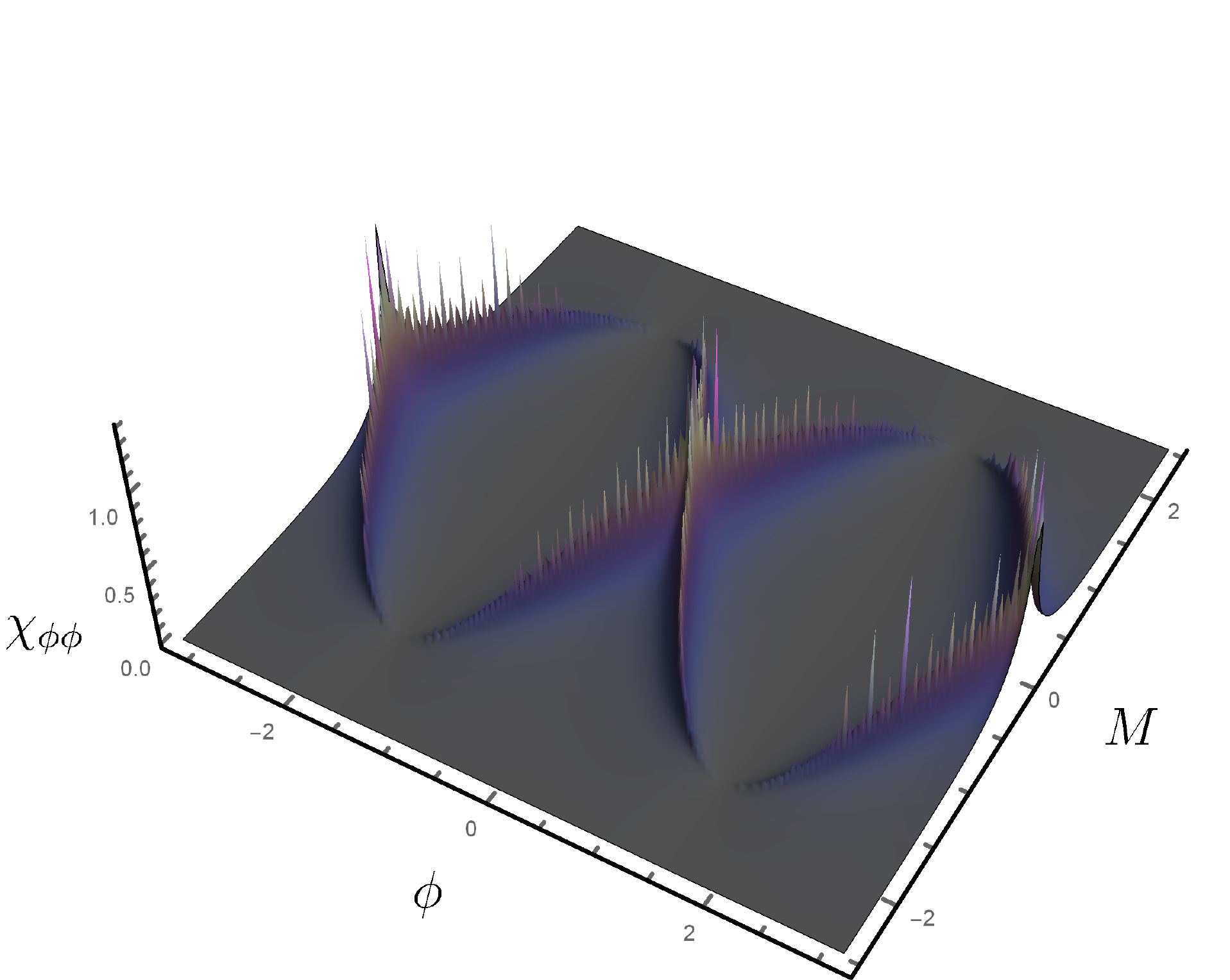}
   \end{subfigure}
   \begin{subfigure}{0.2\textwidth}
   \centering
    \includegraphics[width=\textwidth]{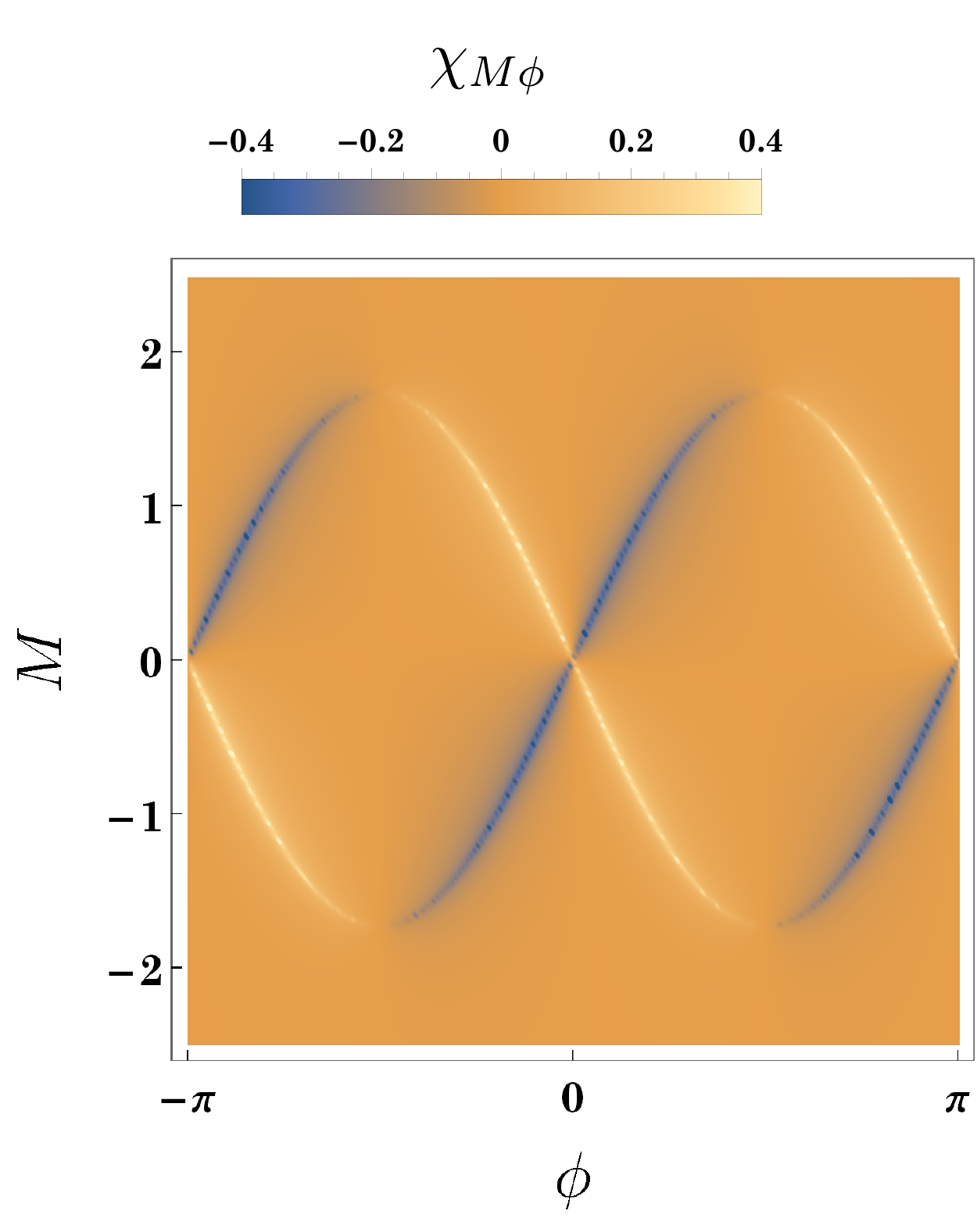}
    \end{subfigure}
    \begin{subfigure}{0.2\textwidth}
    \centering
    \includegraphics[width=\textwidth]{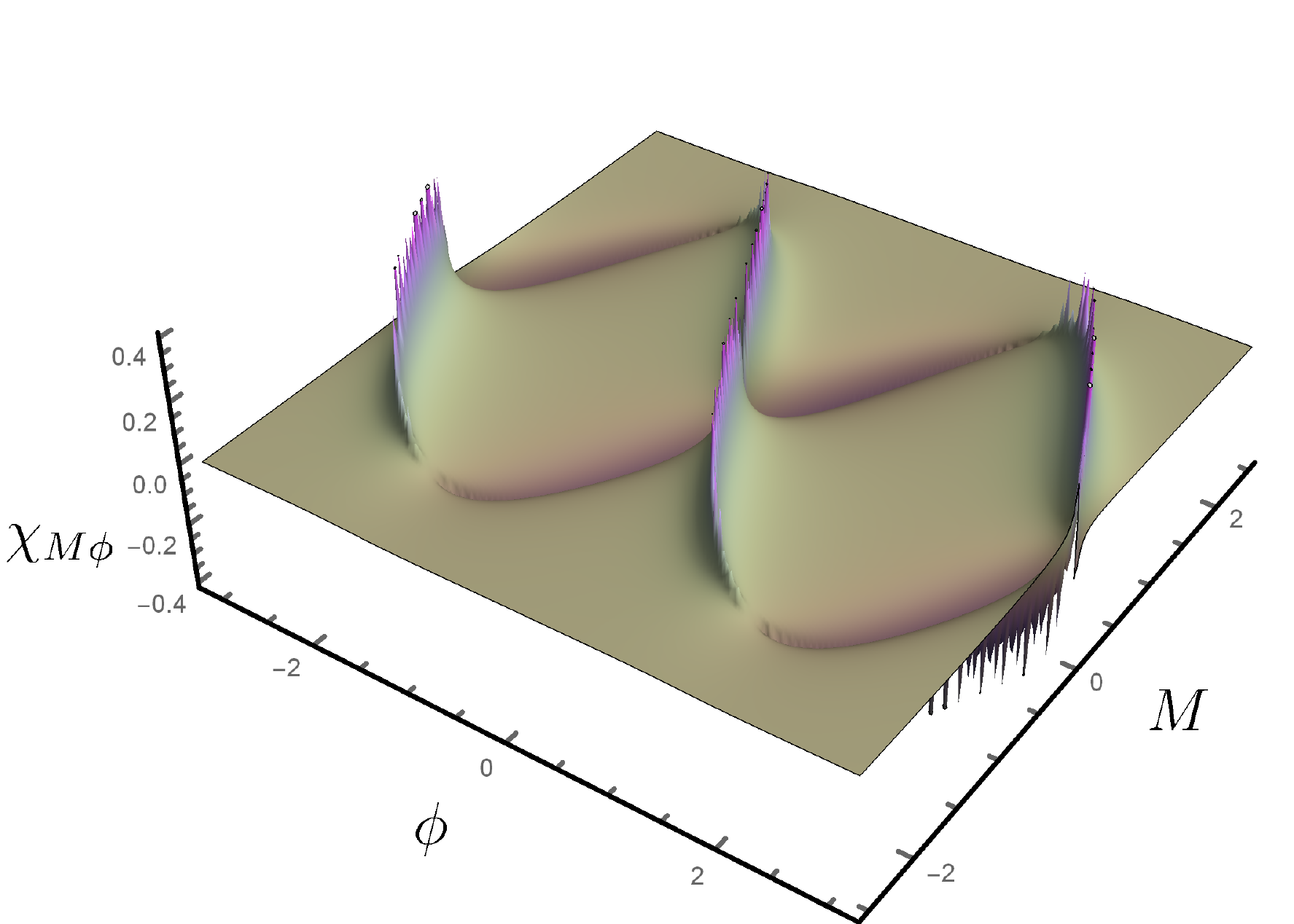}
    \end{subfigure}
    \caption{Fidelity susceptibilities $ \chi_{MM}(\phi,M)$ (top), $\chi_{\phi \phi}(\phi,M)$ (middle), and $\chi_{M\phi}(\phi,M)$ (bottom) calculated at temperature $T = 2 \times 10^{-5}$ and $N_x=N_y=30$ sites.}
    \label{fig:Hald_chi}
\end{figure}

In Fig.~\ref{fig:Hald_chi} we present fidelity susceptibilities (rescaled quantum metric diagonal components) $\chi_{MM} = g_{MM}/L^2$ (the upper plots) and $\chi_{\phi \phi} = g_{\phi\phi}/L^2$ (the middle plots) calculated at temperature $T = 2 \times 10^{-5}$ and $N_x=N_y=30$ sites. We see that both plots clearly show the critical lines, apart from parts of the second plot when the change of the parameter $\phi$ is precisely tangent to the critical manifold, showing that the tangent directions have decreased distinguishability. Just like the case of the XY model, the points $(M,\phi)=(0,0)$ and $(0,\pi)$ behave as isolated critical points, and the metric blows up in all directions in the thermodynamic limit. 

The off-diagonal contribution to the metric shows a very interesting behavior, which we proceed to explain. First, it is consistent with what is inferred from the diagonal contributions, as it also singles out the critical region. Secondly, it gives us additional information about the tangent spaces to the critical submanifold. Namely, we see that $\chi_{M\phi}$ changes sign along the critical region in six points, where it is necessarily zero. The reason for this is twofold. Recall that $\chi_{M\phi}$ is given by the scalar product between $\frac{\partial}{\partial M}$ and $\frac{\partial}{\partial \phi}$. At the ``isolated critical points'' $(M,\phi)=(0,0)$ and $(0,\pi)$, this scalar product becomes zero and the two vectors become orthogonal. On the other hand, at the other four points, characterized by $\phi=\pm \pi/2$, the vector $\frac{\partial}{\partial\phi}$ is precisely tangent to the critical region, which can be inferred from the plot of~$\chi_{\phi\phi}$.

To further analytically confirm the behaviors exhibited in Fig.~\ref{fig:Hald_chi}, one would need to perform careful expansions, as we did in the Appendix for the (modified) XY model. The analysis, however, is similar, but more cumbersome. Since our intent is not to solve particular models, but rather to communicate the idea that tangent directions to critical manifolds are special, we decided to leave these technical details to future work.
\\
\section{Conclusions}
\label{sec: conclusions}
In this paper, we analyzed the thermodynamical limit of the quantum metric along critical submanifolds of theory space. We related its singular behavior to normal directions, which are naturally associated with relevant operators. In the paradigmatic examples of the XY and Haldane models, we have seen (exactly for the first model and numerically for the second) that the normal directions to the critical submanifolds are precisely those where the metric has singular behaviour in the thermodynamical limit, while the tangent ones vanish. In both of these models, the critical regions consist of critical submanifolds, which intersect transversally, and the intersection points behave as isolated critical points with enhanced distinguishability. Further analysis of this phenomenon can serve as a future line of research. As we have seen, our theory predicts that there can also be directions associated to relevant operators that lie in the kernel of $g_*$, a remark which also deserves further investigation. It would be interesting to understand how the above results generalize to the finite-temperature case, where Eq.~\eqref{eq: 2pointcorr and metric} relating the metric and two-point correlation functions gets modified for the case of the Bures metric. Unlike the Bures metric, for the case of the interferometric metric considered in Ref.~\cite{sil:mer:pau:21}, the same functional form as~\eqref{eq: 2pointcorr and metric} is retained---with the expectation value now taken with respect to the appropriate density matrix, and it would be interesting to understand, within the present context, what the physical relevance of considering either metric is. Additionally, the Bogoliubov-Kubo-Mori (BKM) Fisher metric~\cite{pet:07, shi:ued:16} seems to be another relevant physical metric in the context of finite-temperature systems, and it would be interesting to understand this case as well.\\

\section*{Acknowledgments}
\label{sec: acknowledgments}
BM and NP's work was partially supported by SQIG -- Security and Quantum Information Group of Instituto de Telecomunica\c{c}\~oes, by Programme (COMPETE 2020) of the Portugal 2020 framework [Project Q.DOT No.\ 039728 (POCI-01-0247-FEDER-039728)] and the Funda\c{c}\~ao para a Ci\^{e}ncia e a Tecnologia (FCT) through national funds, by FEDER, COMPETE 2020, and by Regional Operational Program of Lisbon, under UIDB/50008/2020 (actions QuRUNNER, QUESTS), Project QuantumMining POCI-01-0145-FEDER-031826 and Project PREDICT PTDC/CCI-CIF/29877/2017.

NP acknowledges FCT Projects CERN/FIS-PAR/0023/2019, QuantumPrime PTDC/EEI-TEL/8017/2020, and the FCT Est\'{i}mulo ao Emprego Cient\'{i}fico Grant No. CEECIND/04594/2017/CP1393/CT000.

This work was also supported through grant No. \ UID/CTM/04540/2019.

\bibliographystyle{apsrev4-1}
\bibliography{bib.bib}

\begin{thebibliography}{53}%
\makeatletter
\providecommand \@ifxundefined [1]{%
 \@ifx{#1\undefined}
}%
\providecommand \@ifnum [1]{%
 \ifnum #1\expandafter \@firstoftwo
 \else \expandafter \@secondoftwo
 \fi
}%
\providecommand \@ifx [1]{%
 \ifx #1\expandafter \@firstoftwo
 \else \expandafter \@secondoftwo
 \fi
}%
\providecommand \natexlab [1]{#1}%
\providecommand \enquote  [1]{``#1''}%
\providecommand \bibnamefont  [1]{#1}%
\providecommand \bibfnamefont [1]{#1}%
\providecommand \citenamefont [1]{#1}%
\providecommand \href@noop [0]{\@secondoftwo}%
\providecommand \href [0]{\begingroup \@sanitize@url \@href}%
\providecommand \@href[1]{\@@startlink{#1}\@@href}%
\providecommand \@@href[1]{\endgroup#1\@@endlink}%
\providecommand \@sanitize@url [0]{\catcode `\\12\catcode `\$12\catcode
  `\&12\catcode `\#12\catcode `\^12\catcode `\_12\catcode `\%12\relax}%
\providecommand \@@startlink[1]{}%
\providecommand \@@endlink[0]{}%
\providecommand \url  [0]{\begingroup\@sanitize@url \@url }%
\providecommand \@url [1]{\endgroup\@href {#1}{\urlprefix }}%
\providecommand \urlprefix  [0]{URL }%
\providecommand \Eprint [0]{\href }%
\providecommand \doibase [0]{http://dx.doi.org/}%
\providecommand \selectlanguage [0]{\@gobble}%
\providecommand \bibinfo  [0]{\@secondoftwo}%
\providecommand \bibfield  [0]{\@secondoftwo}%
\providecommand \translation [1]{[#1]}%
\providecommand \BibitemOpen [0]{}%
\providecommand \bibitemStop [0]{}%
\providecommand \bibitemNoStop [0]{.\EOS\space}%
\providecommand \EOS [0]{\spacefactor3000\relax}%
\providecommand \BibitemShut  [1]{\csname bibitem#1\endcsname}%
\let\auto@bib@innerbib\@empty
\bibitem [{\citenamefont {Amari}\ and\ \citenamefont
  {Nagaoka}(2000)}]{ama:nag:20}%
  \BibitemOpen
  \bibfield  {author} {\bibinfo {author} {\bibfnamefont {S.}~\bibnamefont
  {Amari}}\ and\ \bibinfo {author} {\bibfnamefont {H.}~\bibnamefont
  {Nagaoka}},\ }\href@noop {} {\emph {\bibinfo {title} {{Methods of information
  geometry}}}},\ Vol.\ \bibinfo {volume} {191}\ (\bibinfo  {publisher}
  {American Mathematical Soc.},\ \bibinfo {address} {Providence, Rhode Island,
  U.S.},\ \bibinfo {year} {2000})\BibitemShut {NoStop}%
\bibitem [{\citenamefont {Balian}\ \emph {et~al.}(1986)\citenamefont {Balian},
  \citenamefont {Alhassid},\ and\ \citenamefont {Reinhardt}}]{bal:alh:rei:86}%
  \BibitemOpen
  \bibfield  {author} {\bibinfo {author} {\bibfnamefont {R.}~\bibnamefont
  {Balian}}, \bibinfo {author} {\bibfnamefont {Y.}~\bibnamefont {Alhassid}}, \
  and\ \bibinfo {author} {\bibfnamefont {H.}~\bibnamefont {Reinhardt}},\ }\href
  {\doibase 10.1016/0370-1573(86)90005-0} {\bibfield  {journal} {\bibinfo
  {journal} {Phys. Rept.}\ }\textbf {\bibinfo {volume} {131}},\ \bibinfo
  {pages} {1} (\bibinfo {year} {1986})}\BibitemShut {NoStop}%
\bibitem [{\citenamefont {Mera}\ \emph
  {et~al.}(2022{\natexlab{a}})\citenamefont {Mera}, \citenamefont {Mateus},\
  and\ \citenamefont {Carvalho}}]{mer:mat:car:20}%
  \BibitemOpen
  \bibfield  {author} {\bibinfo {author} {\bibfnamefont {B.}~\bibnamefont
  {Mera}}, \bibinfo {author} {\bibfnamefont {P.}~\bibnamefont {Mateus}}, \ and\
  \bibinfo {author} {\bibfnamefont {A.~M.}\ \bibnamefont {Carvalho}},\ }\href
  {\doibase 10.1109/TIT.2022.3176470} {\bibfield  {journal} {\bibinfo
  {journal} {IEEE Transactions on Information Theory}\ }\textbf {\bibinfo
  {volume} {68}},\ \bibinfo {pages} {5619} (\bibinfo {year}
  {2022}{\natexlab{a}})}\BibitemShut {NoStop}%
\bibitem [{\citenamefont {Zanardi}\ and\ \citenamefont
  {Paunkovi\ifmmode~\acute{c}\else \'{c}\fi{}}(2006)}]{zan:pau:06}%
  \BibitemOpen
  \bibfield  {author} {\bibinfo {author} {\bibfnamefont {P.}~\bibnamefont
  {Zanardi}}\ and\ \bibinfo {author} {\bibfnamefont {N.}~\bibnamefont
  {Paunkovi\ifmmode~\acute{c}\else \'{c}\fi{}}},\ }\href {\doibase
  10.1103/PhysRevE.74.031123} {\bibfield  {journal} {\bibinfo  {journal} {Phys.
  Rev. E}\ }\textbf {\bibinfo {volume} {74}},\ \bibinfo {pages} {031123}
  (\bibinfo {year} {2006})}\BibitemShut {NoStop}%
\bibitem [{\citenamefont {Campos~Venuti}\ and\ \citenamefont
  {Zanardi}(2007)}]{ven:zan:07}%
  \BibitemOpen
  \bibfield  {author} {\bibinfo {author} {\bibfnamefont {L.}~\bibnamefont
  {Campos~Venuti}}\ and\ \bibinfo {author} {\bibfnamefont {P.}~\bibnamefont
  {Zanardi}},\ }\href {\doibase 10.1103/PhysRevLett.99.095701} {\bibfield
  {journal} {\bibinfo  {journal} {Phys. Rev. Lett.}\ }\textbf {\bibinfo
  {volume} {99}},\ \bibinfo {pages} {095701} (\bibinfo {year}
  {2007})}\BibitemShut {NoStop}%
\bibitem [{\citenamefont {Zanardi}\ \emph {et~al.}(2007)\citenamefont
  {Zanardi}, \citenamefont {Campos~Venuti},\ and\ \citenamefont
  {Giorda}}]{zan:ven:07:thermalbures}%
  \BibitemOpen
  \bibfield  {author} {\bibinfo {author} {\bibfnamefont {P.}~\bibnamefont
  {Zanardi}}, \bibinfo {author} {\bibfnamefont {L.}~\bibnamefont
  {Campos~Venuti}}, \ and\ \bibinfo {author} {\bibfnamefont {P.}~\bibnamefont
  {Giorda}},\ }\href {\doibase 10.1103/PhysRevA.76.062318} {\bibfield
  {journal} {\bibinfo  {journal} {Phys. Rev. A}\ }\textbf {\bibinfo {volume}
  {76}},\ \bibinfo {pages} {062318} (\bibinfo {year} {2007})}\BibitemShut
  {NoStop}%
\bibitem [{\citenamefont {Rezakhani}\ \emph {et~al.}(2010)\citenamefont
  {Rezakhani}, \citenamefont {Abasto}, \citenamefont {Lidar},\ and\
  \citenamefont {Zanardi}}]{rez:zan:10}%
  \BibitemOpen
  \bibfield  {author} {\bibinfo {author} {\bibfnamefont {A.~T.}\ \bibnamefont
  {Rezakhani}}, \bibinfo {author} {\bibfnamefont {D.~F.}\ \bibnamefont
  {Abasto}}, \bibinfo {author} {\bibfnamefont {D.~A.}\ \bibnamefont {Lidar}}, \
  and\ \bibinfo {author} {\bibfnamefont {P.}~\bibnamefont {Zanardi}},\ }\href
  {\doibase 10.1103/PhysRevA.82.012321} {\bibfield  {journal} {\bibinfo
  {journal} {Phys. Rev. A}\ }\textbf {\bibinfo {volume} {82}},\ \bibinfo
  {pages} {012321} (\bibinfo {year} {2010})}\BibitemShut {NoStop}%
\bibitem [{\citenamefont {Paunkovi\ifmmode~\acute{c}\else \'{c}\fi{}}\ and\
  \citenamefont {Rocha~Vieira}(2008)}]{pau:vie:08}%
  \BibitemOpen
  \bibfield  {author} {\bibinfo {author} {\bibfnamefont {N.}~\bibnamefont
  {Paunkovi\ifmmode~\acute{c}\else \'{c}\fi{}}}\ and\ \bibinfo {author}
  {\bibfnamefont {V.}~\bibnamefont {Rocha~Vieira}},\ }\href {\doibase
  10.1103/PhysRevE.77.011129} {\bibfield  {journal} {\bibinfo  {journal} {Phys.
  Rev. E}\ }\textbf {\bibinfo {volume} {77}},\ \bibinfo {pages} {011129}
  (\bibinfo {year} {2008})}\BibitemShut {NoStop}%
\bibitem [{\citenamefont {Petz}(2007)}]{pet:07}%
  \BibitemOpen
  \bibfield  {author} {\bibinfo {author} {\bibfnamefont {D.}~\bibnamefont
  {Petz}},\ }\href@noop {} {\emph {\bibinfo {title} {{Quantum information
  theory and quantum statistics}}}}\ (\bibinfo  {publisher} {Springer Science
  \& Business Media},\ \bibinfo {address} {Berlin/Heidelberg, Germany},\
  \bibinfo {year} {2007})\BibitemShut {NoStop}%
\bibitem [{\citenamefont {Bengtsson}\ and\ \citenamefont
  {{\.Z}yczkowski}(2017)}]{ben:zyc:17}%
  \BibitemOpen
  \bibfield  {author} {\bibinfo {author} {\bibfnamefont {I.}~\bibnamefont
  {Bengtsson}}\ and\ \bibinfo {author} {\bibfnamefont {K.}~\bibnamefont
  {{\.Z}yczkowski}},\ }\href@noop {} {\emph {\bibinfo {title} {Geometry of
  quantum states: an introduction to quantum entanglement}}}\ (\bibinfo
  {publisher} {Cambridge university press},\ \bibinfo {address} {Cambridge,
  England},\ \bibinfo {year} {2017})\BibitemShut {NoStop}%
\bibitem [{\citenamefont {Wootters}(1981)}]{woo:81}%
  \BibitemOpen
  \bibfield  {author} {\bibinfo {author} {\bibfnamefont {W.~K.}\ \bibnamefont
  {Wootters}},\ }\href {\doibase 10.1103/PhysRevD.23.357} {\bibfield  {journal}
  {\bibinfo  {journal} {Phys. Rev. D}\ }\textbf {\bibinfo {volume} {23}},\
  \bibinfo {pages} {357} (\bibinfo {year} {1981})}\BibitemShut {NoStop}%
\bibitem [{\citenamefont {Braunstein}\ and\ \citenamefont
  {Caves}(1994)}]{bra:cav:94}%
  \BibitemOpen
  \bibfield  {author} {\bibinfo {author} {\bibfnamefont {S.~L.}\ \bibnamefont
  {Braunstein}}\ and\ \bibinfo {author} {\bibfnamefont {C.~M.}\ \bibnamefont
  {Caves}},\ }\href {\doibase 10.1103/PhysRevLett.72.3439} {\bibfield
  {journal} {\bibinfo  {journal} {Phys. Rev. Lett.}\ }\textbf {\bibinfo
  {volume} {72}},\ \bibinfo {pages} {3439} (\bibinfo {year}
  {1994})}\BibitemShut {NoStop}%
\bibitem [{\citenamefont {Fujiwara}\ and\ \citenamefont
  {Nagaoka}(1995)}]{fuj:nag:95}%
  \BibitemOpen
  \bibfield  {author} {\bibinfo {author} {\bibfnamefont {A.}~\bibnamefont
  {Fujiwara}}\ and\ \bibinfo {author} {\bibfnamefont {H.}~\bibnamefont
  {Nagaoka}},\ }\href {\doibase https://doi.org/10.1016/0375-9601(95)00269-9}
  {\bibfield  {journal} {\bibinfo  {journal} {Physics Letters A}\ }\textbf
  {\bibinfo {volume} {201}},\ \bibinfo {pages} {119} (\bibinfo {year}
  {1995})}\BibitemShut {NoStop}%
\bibitem [{\citenamefont {Brody}\ and\ \citenamefont
  {Hughston}(1998)}]{bro:lan:98}%
  \BibitemOpen
  \bibfield  {author} {\bibinfo {author} {\bibfnamefont {D.~C.}\ \bibnamefont
  {Brody}}\ and\ \bibinfo {author} {\bibfnamefont {L.~P.}\ \bibnamefont
  {Hughston}},\ }\href {\doibase 10.1098/rspa.1998.0266} {\bibfield  {journal}
  {\bibinfo  {journal} {Proceedings of the Royal Society of London. Series A:
  Mathematical, Physical and Engineering Sciences}\ }\textbf {\bibinfo {volume}
  {454}},\ \bibinfo {pages} {2445 } (\bibinfo {year} {1998})}\BibitemShut
  {NoStop}%
\bibitem [{\citenamefont {Liu}\ \emph {et~al.}(2019)\citenamefont {Liu},
  \citenamefont {Yuan}, \citenamefont {Lu},\ and\ \citenamefont
  {Wang}}]{liu:yua:lu:wan:19}%
  \BibitemOpen
  \bibfield  {author} {\bibinfo {author} {\bibfnamefont {J.}~\bibnamefont
  {Liu}}, \bibinfo {author} {\bibfnamefont {H.}~\bibnamefont {Yuan}}, \bibinfo
  {author} {\bibfnamefont {X.-M.}\ \bibnamefont {Lu}}, \ and\ \bibinfo {author}
  {\bibfnamefont {X.}~\bibnamefont {Wang}},\ }\href {\doibase
  10.1088/1751-8121/ab5d4d} {\bibfield  {journal} {\bibinfo  {journal} {Journal
  of Physics A: Mathematical and Theoretical}\ }\textbf {\bibinfo {volume}
  {53}},\ \bibinfo {pages} {023001} (\bibinfo {year} {2019})}\BibitemShut
  {NoStop}%
\bibitem [{\citenamefont {Sidhu}\ and\ \citenamefont
  {Kok}(2020)}]{sid:jas:kok:20}%
  \BibitemOpen
  \bibfield  {author} {\bibinfo {author} {\bibfnamefont {J.~S.}\ \bibnamefont
  {Sidhu}}\ and\ \bibinfo {author} {\bibfnamefont {P.}~\bibnamefont {Kok}},\
  }\href {\doibase 10.1116/1.5119961} {\bibfield  {journal} {\bibinfo
  {journal} {AVS Quantum Science}\ }\textbf {\bibinfo {volume} {2}},\ \bibinfo
  {pages} {014701} (\bibinfo {year} {2020})}\BibitemShut {NoStop}%
\bibitem [{\citenamefont {Suzuki}\ \emph {et~al.}(2020)\citenamefont {Suzuki},
  \citenamefont {Yang},\ and\ \citenamefont {Hayashi}}]{suz:yan:hay:20}%
  \BibitemOpen
  \bibfield  {author} {\bibinfo {author} {\bibfnamefont {J.}~\bibnamefont
  {Suzuki}}, \bibinfo {author} {\bibfnamefont {Y.}~\bibnamefont {Yang}}, \ and\
  \bibinfo {author} {\bibfnamefont {M.}~\bibnamefont {Hayashi}},\ }\href
  {\doibase 10.1088/1751-8121/ab8b78} {\bibfield  {journal} {\bibinfo
  {journal} {Journal of Physics A: Mathematical and Theoretical}\ }\textbf
  {\bibinfo {volume} {53}},\ \bibinfo {pages} {453001} (\bibinfo {year}
  {2020})}\BibitemShut {NoStop}%
\bibitem [{\citenamefont {Provost}\ and\ \citenamefont
  {Vallee}(1980)}]{pro:val:80}%
  \BibitemOpen
  \bibfield  {author} {\bibinfo {author} {\bibfnamefont {J.}~\bibnamefont
  {Provost}}\ and\ \bibinfo {author} {\bibfnamefont {G.}~\bibnamefont
  {Vallee}},\ }\href {\doibase 10.1007/BF02193559} {\bibfield  {journal}
  {\bibinfo  {journal} {Communications in Mathematical Physics}\ }\textbf
  {\bibinfo {volume} {76}},\ \bibinfo {pages} {289} (\bibinfo {year}
  {1980})}\BibitemShut {NoStop}%
\bibitem [{\citenamefont {Mera}\ \emph {et~al.}(2017)\citenamefont {Mera},
  \citenamefont {Vlachou}, \citenamefont {Paunkovi\ifmmode~\acute{c}\else
  \'{c}\fi{}},\ and\ \citenamefont {Vieira}}]{mer:vla:pau:vie:17}%
  \BibitemOpen
  \bibfield  {author} {\bibinfo {author} {\bibfnamefont {B.}~\bibnamefont
  {Mera}}, \bibinfo {author} {\bibfnamefont {C.}~\bibnamefont {Vlachou}},
  \bibinfo {author} {\bibfnamefont {N.}~\bibnamefont
  {Paunkovi\ifmmode~\acute{c}\else \'{c}\fi{}}}, \ and\ \bibinfo {author}
  {\bibfnamefont {V.~R.}\ \bibnamefont {Vieira}},\ }\href {\doibase
  10.1103/PhysRevLett.119.015702} {\bibfield  {journal} {\bibinfo  {journal}
  {Phys. Rev. Lett.}\ }\textbf {\bibinfo {volume} {119}},\ \bibinfo {pages}
  {015702} (\bibinfo {year} {2017})}\BibitemShut {NoStop}%
\bibitem [{\citenamefont {Mera}\ \emph {et~al.}(2018)\citenamefont {Mera},
  \citenamefont {Vlachou}, \citenamefont {Paunkovi\ifmmode~\acute{c}\else
  \'{c}\fi{}}, \citenamefont {Vieira},\ and\ \citenamefont
  {Viyuela}}]{mer:vla:pau:vie:viy:18}%
  \BibitemOpen
  \bibfield  {author} {\bibinfo {author} {\bibfnamefont {B.}~\bibnamefont
  {Mera}}, \bibinfo {author} {\bibfnamefont {C.}~\bibnamefont {Vlachou}},
  \bibinfo {author} {\bibfnamefont {N.}~\bibnamefont
  {Paunkovi\ifmmode~\acute{c}\else \'{c}\fi{}}}, \bibinfo {author}
  {\bibfnamefont {V.~R.}\ \bibnamefont {Vieira}}, \ and\ \bibinfo {author}
  {\bibfnamefont {O.}~\bibnamefont {Viyuela}},\ }\href {\doibase
  10.1103/PhysRevB.97.094110} {\bibfield  {journal} {\bibinfo  {journal} {Phys.
  Rev. B}\ }\textbf {\bibinfo {volume} {97}},\ \bibinfo {pages} {094110}
  (\bibinfo {year} {2018})}\BibitemShut {NoStop}%
\bibitem [{\citenamefont {Amin}\ \emph {et~al.}(2018)\citenamefont {Amin},
  \citenamefont {Mera}, \citenamefont {Vlachou}, \citenamefont
  {Paunkovi\ifmmode~\acute{c}\else \'{c}\fi{}},\ and\ \citenamefont
  {Vieira}}]{ami:mer:vla:pau:vie:18}%
  \BibitemOpen
  \bibfield  {author} {\bibinfo {author} {\bibfnamefont {S.~T.}\ \bibnamefont
  {Amin}}, \bibinfo {author} {\bibfnamefont {B.}~\bibnamefont {Mera}}, \bibinfo
  {author} {\bibfnamefont {C.}~\bibnamefont {Vlachou}}, \bibinfo {author}
  {\bibfnamefont {N.}~\bibnamefont {Paunkovi\ifmmode~\acute{c}\else
  \'{c}\fi{}}}, \ and\ \bibinfo {author} {\bibfnamefont {V.~R.}\ \bibnamefont
  {Vieira}},\ }\href {\doibase 10.1103/PhysRevB.98.245141} {\bibfield
  {journal} {\bibinfo  {journal} {Phys. Rev. B}\ }\textbf {\bibinfo {volume}
  {98}},\ \bibinfo {pages} {245141} (\bibinfo {year} {2018})}\BibitemShut
  {NoStop}%
\bibitem [{\citenamefont {Peotta}\ and\ \citenamefont
  {T{\"o}rm{\"a}}(2015)}]{peotta2015superfluidity}%
  \BibitemOpen
  \bibfield  {author} {\bibinfo {author} {\bibfnamefont {S.}~\bibnamefont
  {Peotta}}\ and\ \bibinfo {author} {\bibfnamefont {P.}~\bibnamefont
  {T{\"o}rm{\"a}}},\ }\href {\doibase 10.1038/ncomms9944} {\bibfield  {journal}
  {\bibinfo  {journal} {Nature communications}\ }\textbf {\bibinfo {volume}
  {6}},\ \bibinfo {pages} {1} (\bibinfo {year} {2015})}\BibitemShut {NoStop}%
\bibitem [{\citenamefont {Julku}\ \emph {et~al.}(2016)\citenamefont {Julku},
  \citenamefont {Peotta}, \citenamefont {Vanhala}, \citenamefont {Kim},\ and\
  \citenamefont {T\"orm\"a}}]{julku2016geometric}%
  \BibitemOpen
  \bibfield  {author} {\bibinfo {author} {\bibfnamefont {A.}~\bibnamefont
  {Julku}}, \bibinfo {author} {\bibfnamefont {S.}~\bibnamefont {Peotta}},
  \bibinfo {author} {\bibfnamefont {T.~I.}\ \bibnamefont {Vanhala}}, \bibinfo
  {author} {\bibfnamefont {D.-H.}\ \bibnamefont {Kim}}, \ and\ \bibinfo
  {author} {\bibfnamefont {P.}~\bibnamefont {T\"orm\"a}},\ }\href {\doibase
  10.1103/PhysRevLett.117.045303} {\bibfield  {journal} {\bibinfo  {journal}
  {Phys. Rev. Lett.}\ }\textbf {\bibinfo {volume} {117}},\ \bibinfo {pages}
  {045303} (\bibinfo {year} {2016})}\BibitemShut {NoStop}%
\bibitem [{\citenamefont {Liang}\ \emph {et~al.}(2017)\citenamefont {Liang},
  \citenamefont {Vanhala}, \citenamefont {Peotta}, \citenamefont {Siro},
  \citenamefont {Harju},\ and\ \citenamefont {T\"orm\"a}}]{liang2017band}%
  \BibitemOpen
  \bibfield  {author} {\bibinfo {author} {\bibfnamefont {L.}~\bibnamefont
  {Liang}}, \bibinfo {author} {\bibfnamefont {T.~I.}\ \bibnamefont {Vanhala}},
  \bibinfo {author} {\bibfnamefont {S.}~\bibnamefont {Peotta}}, \bibinfo
  {author} {\bibfnamefont {T.}~\bibnamefont {Siro}}, \bibinfo {author}
  {\bibfnamefont {A.}~\bibnamefont {Harju}}, \ and\ \bibinfo {author}
  {\bibfnamefont {P.}~\bibnamefont {T\"orm\"a}},\ }\href {\doibase
  10.1103/PhysRevB.95.024515} {\bibfield  {journal} {\bibinfo  {journal} {Phys.
  Rev. B}\ }\textbf {\bibinfo {volume} {95}},\ \bibinfo {pages} {024515}
  (\bibinfo {year} {2017})}\BibitemShut {NoStop}%
\bibitem [{\citenamefont {Iskin}(2018)}]{iskin2018quantum}%
  \BibitemOpen
  \bibfield  {author} {\bibinfo {author} {\bibfnamefont {M.}~\bibnamefont
  {Iskin}},\ }\href {\doibase 10.1103/PhysRevA.97.033625} {\bibfield  {journal}
  {\bibinfo  {journal} {Phys. Rev. A}\ }\textbf {\bibinfo {volume} {97}},\
  \bibinfo {pages} {033625} (\bibinfo {year} {2018})}\BibitemShut {NoStop}%
\bibitem [{\citenamefont {Parameswaran}\ \emph {et~al.}(2013)\citenamefont
  {Parameswaran}, \citenamefont {Roy},\ and\ \citenamefont
  {Sondhi}}]{par:roy:son:13}%
  \BibitemOpen
  \bibfield  {author} {\bibinfo {author} {\bibfnamefont {S.~A.}\ \bibnamefont
  {Parameswaran}}, \bibinfo {author} {\bibfnamefont {R.}~\bibnamefont {Roy}}, \
  and\ \bibinfo {author} {\bibfnamefont {S.~L.}\ \bibnamefont {Sondhi}},\
  }\href {\doibase 10.1016/j.crhy.2013.04.003} {\bibfield  {journal} {\bibinfo
  {journal} {Comptes Rendus Physique}\ }\textbf {\bibinfo {volume} {14}},\
  \bibinfo {pages} {816} (\bibinfo {year} {2013})}\BibitemShut {NoStop}%
\bibitem [{\citenamefont {Roy}(2014)}]{roy:14}%
  \BibitemOpen
  \bibfield  {author} {\bibinfo {author} {\bibfnamefont {R.}~\bibnamefont
  {Roy}},\ }\href {\doibase 10.1103/PhysRevB.90.165139} {\bibfield  {journal}
  {\bibinfo  {journal} {Phys. Rev. B}\ }\textbf {\bibinfo {volume} {90}},\
  \bibinfo {pages} {165139} (\bibinfo {year} {2014})}\BibitemShut {NoStop}%
\bibitem [{\citenamefont {Claassen}\ \emph {et~al.}(2015)\citenamefont
  {Claassen}, \citenamefont {Lee}, \citenamefont {Thomale}, \citenamefont
  {Qi},\ and\ \citenamefont {Devereaux}}]{cla:lee:tho:qi:dev:15}%
  \BibitemOpen
  \bibfield  {author} {\bibinfo {author} {\bibfnamefont {M.}~\bibnamefont
  {Claassen}}, \bibinfo {author} {\bibfnamefont {C.~H.}\ \bibnamefont {Lee}},
  \bibinfo {author} {\bibfnamefont {R.}~\bibnamefont {Thomale}}, \bibinfo
  {author} {\bibfnamefont {X.-L.}\ \bibnamefont {Qi}}, \ and\ \bibinfo {author}
  {\bibfnamefont {T.~P.}\ \bibnamefont {Devereaux}},\ }\href {\doibase
  10.1103/PhysRevLett.114.236802} {\bibfield  {journal} {\bibinfo  {journal}
  {Phys. Rev. Lett.}\ }\textbf {\bibinfo {volume} {114}},\ \bibinfo {pages}
  {236802} (\bibinfo {year} {2015})}\BibitemShut {NoStop}%
\bibitem [{\citenamefont {Jackson}\ \emph {et~al.}(2015)\citenamefont
  {Jackson}, \citenamefont {M{\"o}ller},\ and\ \citenamefont
  {Roy}}]{jackson2015geometric}%
  \BibitemOpen
  \bibfield  {author} {\bibinfo {author} {\bibfnamefont {T.~S.}\ \bibnamefont
  {Jackson}}, \bibinfo {author} {\bibfnamefont {G.}~\bibnamefont {M{\"o}ller}},
  \ and\ \bibinfo {author} {\bibfnamefont {R.}~\bibnamefont {Roy}},\ }\href
  {\doibase 10.1038/ncomms9629} {\bibfield  {journal} {\bibinfo  {journal}
  {Nature communications}\ }\textbf {\bibinfo {volume} {6}},\ \bibinfo {pages}
  {1} (\bibinfo {year} {2015})}\BibitemShut {NoStop}%
\bibitem [{\citenamefont {Lee}\ \emph {et~al.}(2017)\citenamefont {Lee},
  \citenamefont {Claassen},\ and\ \citenamefont {Thomale}}]{lee:cla:tho:17}%
  \BibitemOpen
  \bibfield  {author} {\bibinfo {author} {\bibfnamefont {C.~H.}\ \bibnamefont
  {Lee}}, \bibinfo {author} {\bibfnamefont {M.}~\bibnamefont {Claassen}}, \
  and\ \bibinfo {author} {\bibfnamefont {R.}~\bibnamefont {Thomale}},\ }\href
  {\doibase 10.1103/PhysRevB.96.165150} {\bibfield  {journal} {\bibinfo
  {journal} {Phys. Rev. B}\ }\textbf {\bibinfo {volume} {96}},\ \bibinfo
  {pages} {165150} (\bibinfo {year} {2017})}\BibitemShut {NoStop}%
\bibitem [{\citenamefont {Wang}\ \emph {et~al.}(2021)\citenamefont {Wang},
  \citenamefont {Cano}, \citenamefont {Millis}, \citenamefont {Liu},\ and\
  \citenamefont {Yang}}]{wang2021exact}%
  \BibitemOpen
  \bibfield  {author} {\bibinfo {author} {\bibfnamefont {J.}~\bibnamefont
  {Wang}}, \bibinfo {author} {\bibfnamefont {J.}~\bibnamefont {Cano}}, \bibinfo
  {author} {\bibfnamefont {A.~J.}\ \bibnamefont {Millis}}, \bibinfo {author}
  {\bibfnamefont {Z.}~\bibnamefont {Liu}}, \ and\ \bibinfo {author}
  {\bibfnamefont {B.}~\bibnamefont {Yang}},\ }\href {\doibase
  10.1103/PhysRevLett.127.246403} {\bibfield  {journal} {\bibinfo  {journal}
  {Phys. Rev. Lett.}\ }\textbf {\bibinfo {volume} {127}},\ \bibinfo {pages}
  {246403} (\bibinfo {year} {2021})}\BibitemShut {NoStop}%
\bibitem [{\citenamefont {Topp}\ \emph {et~al.}(2021)\citenamefont {Topp},
  \citenamefont {Eckhardt}, \citenamefont {Kennes}, \citenamefont {Sentef},\
  and\ \citenamefont {T\"orm\"a}}]{topp2021light}%
  \BibitemOpen
  \bibfield  {author} {\bibinfo {author} {\bibfnamefont {G.~E.}\ \bibnamefont
  {Topp}}, \bibinfo {author} {\bibfnamefont {C.~J.}\ \bibnamefont {Eckhardt}},
  \bibinfo {author} {\bibfnamefont {D.~M.}\ \bibnamefont {Kennes}}, \bibinfo
  {author} {\bibfnamefont {M.~A.}\ \bibnamefont {Sentef}}, \ and\ \bibinfo
  {author} {\bibfnamefont {P.}~\bibnamefont {T\"orm\"a}},\ }\href {\doibase
  10.1103/PhysRevB.104.064306} {\bibfield  {journal} {\bibinfo  {journal}
  {Phys. Rev. B}\ }\textbf {\bibinfo {volume} {104}},\ \bibinfo {pages}
  {064306} (\bibinfo {year} {2021})}\BibitemShut {NoStop}%
\bibitem [{\citenamefont {Marzari}\ and\ \citenamefont
  {Vanderbilt}(1997)}]{mar:97}%
  \BibitemOpen
  \bibfield  {author} {\bibinfo {author} {\bibfnamefont {N.}~\bibnamefont
  {Marzari}}\ and\ \bibinfo {author} {\bibfnamefont {D.}~\bibnamefont
  {Vanderbilt}},\ }\href {\doibase 10.1103/PhysRevB.56.12847} {\bibfield
  {journal} {\bibinfo  {journal} {Phys. Rev. B}\ }\textbf {\bibinfo {volume}
  {56}},\ \bibinfo {pages} {12847} (\bibinfo {year} {1997})}\BibitemShut
  {NoStop}%
\bibitem [{\citenamefont {Ozawa}\ and\ \citenamefont
  {Goldman}(2018)}]{oza:gol:18}%
  \BibitemOpen
  \bibfield  {author} {\bibinfo {author} {\bibfnamefont {T.}~\bibnamefont
  {Ozawa}}\ and\ \bibinfo {author} {\bibfnamefont {N.}~\bibnamefont
  {Goldman}},\ }\href {\doibase 10.1103/PhysRevB.97.201117} {\bibfield
  {journal} {\bibinfo  {journal} {Phys. Rev. B}\ }\textbf {\bibinfo {volume}
  {97}},\ \bibinfo {pages} {201117} (\bibinfo {year} {2018})}\BibitemShut
  {NoStop}%
\bibitem [{\citenamefont {Palumbo}\ and\ \citenamefont
  {Goldman}(2019)}]{pal:gol:19}%
  \BibitemOpen
  \bibfield  {author} {\bibinfo {author} {\bibfnamefont {G.}~\bibnamefont
  {Palumbo}}\ and\ \bibinfo {author} {\bibfnamefont {N.}~\bibnamefont
  {Goldman}},\ }\href {\doibase 10.1103/PhysRevB.99.045154} {\bibfield
  {journal} {\bibinfo  {journal} {Phys. Rev. B}\ }\textbf {\bibinfo {volume}
  {99}},\ \bibinfo {pages} {045154} (\bibinfo {year} {2019})}\BibitemShut
  {NoStop}%
\bibitem [{\citenamefont {Salerno}\ \emph {et~al.}(2020)\citenamefont
  {Salerno}, \citenamefont {Goldman},\ and\ \citenamefont
  {Palumbo}}]{sal:gol:pal:20}%
  \BibitemOpen
  \bibfield  {author} {\bibinfo {author} {\bibfnamefont {G.}~\bibnamefont
  {Salerno}}, \bibinfo {author} {\bibfnamefont {N.}~\bibnamefont {Goldman}}, \
  and\ \bibinfo {author} {\bibfnamefont {G.}~\bibnamefont {Palumbo}},\ }\href
  {\doibase 10.1103/PhysRevResearch.2.013224} {\bibfield  {journal} {\bibinfo
  {journal} {Phys. Rev. Res.}\ }\textbf {\bibinfo {volume} {2}},\ \bibinfo
  {pages} {013224} (\bibinfo {year} {2020})}\BibitemShut {NoStop}%
\bibitem [{\citenamefont {Ozawa}\ and\ \citenamefont
  {Goldman}(2019)}]{oza:gol:19}%
  \BibitemOpen
  \bibfield  {author} {\bibinfo {author} {\bibfnamefont {T.}~\bibnamefont
  {Ozawa}}\ and\ \bibinfo {author} {\bibfnamefont {N.}~\bibnamefont
  {Goldman}},\ }\href {\doibase 10.1103/PhysRevResearch.1.032019} {\bibfield
  {journal} {\bibinfo  {journal} {Phys. Rev. Res.}\ }\textbf {\bibinfo {volume}
  {1}},\ \bibinfo {pages} {032019} (\bibinfo {year} {2019})}\BibitemShut
  {NoStop}%
\bibitem [{\citenamefont {Mera}(2020)}]{mer:20}%
  \BibitemOpen
  \bibfield  {author} {\bibinfo {author} {\bibfnamefont {B.}~\bibnamefont
  {Mera}},\ }\href {\doibase 10.1103/PhysRevB.101.115128} {\bibfield  {journal}
  {\bibinfo  {journal} {Phys. Rev. B}\ }\textbf {\bibinfo {volume} {101}},\
  \bibinfo {pages} {115128} (\bibinfo {year} {2020})}\BibitemShut {NoStop}%
\bibitem [{\citenamefont {Ozawa}\ and\ \citenamefont
  {Mera}(2021)}]{oza:mer:21}%
  \BibitemOpen
  \bibfield  {author} {\bibinfo {author} {\bibfnamefont {T.}~\bibnamefont
  {Ozawa}}\ and\ \bibinfo {author} {\bibfnamefont {B.}~\bibnamefont {Mera}},\
  }\href {\doibase 10.1103/PhysRevB.104.045103} {\bibfield  {journal} {\bibinfo
   {journal} {Phys. Rev. B}\ }\textbf {\bibinfo {volume} {104}},\ \bibinfo
  {pages} {045103} (\bibinfo {year} {2021})}\BibitemShut {NoStop}%
\bibitem [{\citenamefont {Mera}\ and\ \citenamefont
  {Ozawa}(2021{\natexlab{a}})}]{mer:oza:21}%
  \BibitemOpen
  \bibfield  {author} {\bibinfo {author} {\bibfnamefont {B.}~\bibnamefont
  {Mera}}\ and\ \bibinfo {author} {\bibfnamefont {T.}~\bibnamefont {Ozawa}},\
  }\href {\doibase 10.1103/PhysRevB.104.045104} {\bibfield  {journal} {\bibinfo
   {journal} {Phys. Rev. B}\ }\textbf {\bibinfo {volume} {104}},\ \bibinfo
  {pages} {045104} (\bibinfo {year} {2021}{\natexlab{a}})}\BibitemShut
  {NoStop}%
\bibitem [{\citenamefont {Mera}\ and\ \citenamefont
  {Ozawa}(2021{\natexlab{b}})}]{mer:tom:21:engineering}%
  \BibitemOpen
  \bibfield  {author} {\bibinfo {author} {\bibfnamefont {B.}~\bibnamefont
  {Mera}}\ and\ \bibinfo {author} {\bibfnamefont {T.}~\bibnamefont {Ozawa}},\
  }\href {\doibase 10.1103/PhysRevB.104.115160} {\bibfield  {journal} {\bibinfo
   {journal} {Phys. Rev. B}\ }\textbf {\bibinfo {volume} {104}},\ \bibinfo
  {pages} {115160} (\bibinfo {year} {2021}{\natexlab{b}})}\BibitemShut
  {NoStop}%
\bibitem [{\citenamefont {Wang}\ and\ \citenamefont {Liu}(2022)}]{wan:liu:21}%
  \BibitemOpen
  \bibfield  {author} {\bibinfo {author} {\bibfnamefont {J.}~\bibnamefont
  {Wang}}\ and\ \bibinfo {author} {\bibfnamefont {Z.}~\bibnamefont {Liu}},\
  }\href {\doibase 10.1103/PhysRevLett.128.176403} {\bibfield  {journal}
  {\bibinfo  {journal} {Phys. Rev. Lett.}\ }\textbf {\bibinfo {volume} {128}},\
  \bibinfo {pages} {176403} (\bibinfo {year} {2022})}\BibitemShut {NoStop}%
\bibitem [{\citenamefont {Ledwith}\ \emph {et~al.}(2022)\citenamefont
  {Ledwith}, \citenamefont {Vishwanath},\ and\ \citenamefont
  {Khalaf}}]{led:vish:kha:21}%
  \BibitemOpen
  \bibfield  {author} {\bibinfo {author} {\bibfnamefont {P.~J.}\ \bibnamefont
  {Ledwith}}, \bibinfo {author} {\bibfnamefont {A.}~\bibnamefont {Vishwanath}},
  \ and\ \bibinfo {author} {\bibfnamefont {E.}~\bibnamefont {Khalaf}},\ }\href
  {\doibase 10.1103/PhysRevLett.128.176404} {\bibfield  {journal} {\bibinfo
  {journal} {Phys. Rev. Lett.}\ }\textbf {\bibinfo {volume} {128}},\ \bibinfo
  {pages} {176404} (\bibinfo {year} {2022})}\BibitemShut {NoStop}%
\bibitem [{\citenamefont {{T{\"o}rm{\"a}}}\ \emph {et~al.}(2022)\citenamefont
  {{T{\"o}rm{\"a}}}, \citenamefont {{Peotta}},\ and\ \citenamefont
  {{Bernevig}}}]{tor:peo:bern:21}%
  \BibitemOpen
  \bibfield  {author} {\bibinfo {author} {\bibfnamefont {P.}~\bibnamefont
  {{T{\"o}rm{\"a}}}}, \bibinfo {author} {\bibfnamefont {S.}~\bibnamefont
  {{Peotta}}}, \ and\ \bibinfo {author} {\bibfnamefont {B.~A.}\ \bibnamefont
  {{Bernevig}}},\ }\href {\doibase 10.1038/s42254-022-00466-y} {\bibfield
  {journal} {\bibinfo  {journal} {Nature Reviews Physics}\ }\textbf {\bibinfo
  {volume} {4}},\ \bibinfo {pages} {528} (\bibinfo {year} {2022})}\BibitemShut
  {NoStop}%
\bibitem [{\citenamefont {Parker}\ \emph {et~al.}(2021)\citenamefont {Parker},
  \citenamefont {Ledwith}, \citenamefont {Khalaf}, \citenamefont {Soejima},
  \citenamefont {Hauschild}, \citenamefont {Xie}, \citenamefont {Pierce},
  \citenamefont {Zaletel}, \citenamefont {Yacoby},\ and\ \citenamefont
  {Vishwanath}}]{par:led:kha:etal:21}%
  \BibitemOpen
  \bibfield  {author} {\bibinfo {author} {\bibfnamefont {D.}~\bibnamefont
  {Parker}}, \bibinfo {author} {\bibfnamefont {P.}~\bibnamefont {Ledwith}},
  \bibinfo {author} {\bibfnamefont {E.}~\bibnamefont {Khalaf}}, \bibinfo
  {author} {\bibfnamefont {T.}~\bibnamefont {Soejima}}, \bibinfo {author}
  {\bibfnamefont {J.}~\bibnamefont {Hauschild}}, \bibinfo {author}
  {\bibfnamefont {Y.}~\bibnamefont {Xie}}, \bibinfo {author} {\bibfnamefont
  {A.}~\bibnamefont {Pierce}}, \bibinfo {author} {\bibfnamefont {M.~P.}\
  \bibnamefont {Zaletel}}, \bibinfo {author} {\bibfnamefont {A.}~\bibnamefont
  {Yacoby}}, \ and\ \bibinfo {author} {\bibfnamefont {A.}~\bibnamefont
  {Vishwanath}},\ }\href {\doibase 10.48550/arXiv.2112.13837} {\bibfield
  {journal} {\bibinfo  {journal} {arXiv preprint arXiv:2112.13837}\ } (\bibinfo
  {year} {2021}),\ 10.48550/arXiv.2112.13837}\BibitemShut {NoStop}%
\bibitem [{\citenamefont {Northe}\ \emph {et~al.}(2022)\citenamefont {Northe},
  \citenamefont {Palumbo}, \citenamefont {Sturm}, \citenamefont {Tutschku},\
  and\ \citenamefont {Hankiewicz}}]{nor:pal:sturm:21}%
  \BibitemOpen
  \bibfield  {author} {\bibinfo {author} {\bibfnamefont {C.}~\bibnamefont
  {Northe}}, \bibinfo {author} {\bibfnamefont {G.}~\bibnamefont {Palumbo}},
  \bibinfo {author} {\bibfnamefont {J.}~\bibnamefont {Sturm}}, \bibinfo
  {author} {\bibfnamefont {C.}~\bibnamefont {Tutschku}}, \ and\ \bibinfo
  {author} {\bibfnamefont {E.~M.}\ \bibnamefont {Hankiewicz}},\ }\href
  {\doibase 10.1103/PhysRevB.105.155410} {\bibfield  {journal} {\bibinfo
  {journal} {Phys. Rev. B}\ }\textbf {\bibinfo {volume} {105}},\ \bibinfo
  {pages} {155410} (\bibinfo {year} {2022})}\BibitemShut {NoStop}%
\bibitem [{\citenamefont {Mera}\ \emph
  {et~al.}(2022{\natexlab{b}})\citenamefont {Mera}, \citenamefont {Zhang},\
  and\ \citenamefont {Goldman}}]{mer:zha:gol:21}%
  \BibitemOpen
  \bibfield  {author} {\bibinfo {author} {\bibfnamefont {B.}~\bibnamefont
  {Mera}}, \bibinfo {author} {\bibfnamefont {A.}~\bibnamefont {Zhang}}, \ and\
  \bibinfo {author} {\bibfnamefont {N.}~\bibnamefont {Goldman}},\ }\href
  {\doibase 10.21468/SciPostPhys.12.1.018} {\bibfield  {journal} {\bibinfo
  {journal} {SciPost Phys.}\ }\textbf {\bibinfo {volume} {12}},\ \bibinfo
  {pages} {018} (\bibinfo {year} {2022}{\natexlab{b}})}\BibitemShut {NoStop}%
\bibitem [{\citenamefont {Tong}(2017)}]{ton:17}%
  \BibitemOpen
  \bibfield  {author} {\bibinfo {author} {\bibfnamefont {D.}~\bibnamefont
  {Tong}},\ }\href {http://www.damtp.cam.ac.uk/user/tong/sft/sft.pdf} {\enquote
  {\bibinfo {title} {{Lectures on Statistical Field Theory}},}\ } (\bibinfo
  {year} {2017})\BibitemShut {NoStop}%
\bibitem [{\citenamefont {Miyaji}\ \emph {et~al.}(2015)\citenamefont {Miyaji},
  \citenamefont {Numasawa}, \citenamefont {Shiba}, \citenamefont {Takayanagi},\
  and\ \citenamefont {Watanabe}}]{miy:15}%
  \BibitemOpen
  \bibfield  {author} {\bibinfo {author} {\bibfnamefont {M.}~\bibnamefont
  {Miyaji}}, \bibinfo {author} {\bibfnamefont {T.}~\bibnamefont {Numasawa}},
  \bibinfo {author} {\bibfnamefont {N.}~\bibnamefont {Shiba}}, \bibinfo
  {author} {\bibfnamefont {T.}~\bibnamefont {Takayanagi}}, \ and\ \bibinfo
  {author} {\bibfnamefont {K.}~\bibnamefont {Watanabe}},\ }\href {\doibase
  10.1103/PhysRevLett.115.261602} {\bibfield  {journal} {\bibinfo  {journal}
  {Phys. Rev. Lett.}\ }\textbf {\bibinfo {volume} {115}},\ \bibinfo {pages}
  {261602} (\bibinfo {year} {2015})}\BibitemShut {NoStop}%
\bibitem [{\citenamefont {Araki}\ and\ \citenamefont
  {Matsui}(1985)}]{ara:mat:85}%
  \BibitemOpen
  \bibfield  {author} {\bibinfo {author} {\bibfnamefont {H.}~\bibnamefont
  {Araki}}\ and\ \bibinfo {author} {\bibfnamefont {T.}~\bibnamefont {Matsui}},\
  }\href {\doibase 10.1007/BF01218760} {\bibfield  {journal} {\bibinfo
  {journal} {Communications in mathematical physics}\ }\textbf {\bibinfo
  {volume} {101}},\ \bibinfo {pages} {213} (\bibinfo {year}
  {1985})}\BibitemShut {NoStop}%
\bibitem [{\citenamefont {Haldane}(1988)}]{hal:88}%
  \BibitemOpen
  \bibfield  {author} {\bibinfo {author} {\bibfnamefont {F.~D.~M.}\
  \bibnamefont {Haldane}},\ }\href {\doibase 10.1103/PhysRevLett.61.2015}
  {\bibfield  {journal} {\bibinfo  {journal} {Phys. Rev. Lett.}\ }\textbf
  {\bibinfo {volume} {61}},\ \bibinfo {pages} {2015} (\bibinfo {year}
  {1988})}\BibitemShut {NoStop}%
\bibitem [{\citenamefont {Silva}\ \emph {et~al.}(2021)\citenamefont {Silva},
  \citenamefont {Mera},\ and\ \citenamefont {Paunkovi\ifmmode~\acute{c}\else
  \'{c}\fi{}}}]{sil:mer:pau:21}%
  \BibitemOpen
  \bibfield  {author} {\bibinfo {author} {\bibfnamefont {H.}~\bibnamefont
  {Silva}}, \bibinfo {author} {\bibfnamefont {B.}~\bibnamefont {Mera}}, \ and\
  \bibinfo {author} {\bibfnamefont {N.}~\bibnamefont
  {Paunkovi\ifmmode~\acute{c}\else \'{c}\fi{}}},\ }\href {\doibase
  10.1103/PhysRevB.103.085127} {\bibfield  {journal} {\bibinfo  {journal}
  {Phys. Rev. B}\ }\textbf {\bibinfo {volume} {103}},\ \bibinfo {pages}
  {085127} (\bibinfo {year} {2021})}\BibitemShut {NoStop}%
\bibitem [{\citenamefont {Shitara}\ and\ \citenamefont
  {Ueda}(2016)}]{shi:ued:16}%
  \BibitemOpen
  \bibfield  {author} {\bibinfo {author} {\bibfnamefont {T.}~\bibnamefont
  {Shitara}}\ and\ \bibinfo {author} {\bibfnamefont {M.}~\bibnamefont {Ueda}},\
  }\href {\doibase 10.1103/PhysRevA.94.062316} {\bibfield  {journal} {\bibinfo
  {journal} {Phys. Rev. A}\ }\textbf {\bibinfo {volume} {94}},\ \bibinfo
  {pages} {062316} (\bibinfo {year} {2016})}\BibitemShut {NoStop}%
\end{thebibliography}%

\appendix 
\begin{widetext}
\section{Thermodynamic limit of the quantum metric}
\label{appendix: td limit of g}
In this appendix, we analyze the thermodynamical limit of the quantum metric in the case of the anisotropic XY spin chain in an external magnetic field. A straightforward calculation shows that the components of the quantum metric are given by
\begin{align*}
g_{\lambda\lambda} &=\frac{1}{4}\sum_{k}\frac{\gamma^2\sin^2(k)}{\left[\left(\lambda -\cos(k)\right)^2 +\gamma^2 \sin^2k\right]^2},
\end{align*}
\begin{align*}
g_{\gamma\gamma} &=\frac{1}{4}\sum_{k}\frac{\left(\lambda-\cos(k)\right)^2\sin^2(k)}{\left[\left(\lambda -\cos(k)\right)^2 +\gamma^2 \sin^2k\right]^2},
\end{align*}
\begin{align*}
g_{\lambda \gamma}=\frac{1}{4}\sum_{k}\frac{\gamma  \sin ^2(k) (\cos (k)-\lambda )}{\left(\gamma ^2 \sin ^2(k)+(\lambda -\cos (k))^2\right)^2}.    
\end{align*}

Below, we analyze the large-$N$ behaviour of the above expressions in the critical region.

\subsection{Segment $\gamma=0$ and $|\lambda| \leq 1$}
For the critical submanifold defined by $\gamma=0$ and $|\lambda| \leq 1$, with tangent vector $\frac{\partial}{\partial \lambda}$, we immediately see that $g_{\lambda\lambda}=0$ independently of $N$. For $g_{\gamma\gamma}$ the metric becomes
\begin{align*}
g_{\gamma\gamma}(\lambda,0)=\frac{1}{4}\sum_{k}\frac{\sin^2(k)}{\left(\lambda -\cos(k)\right)^2 }.
\end{align*}

For $g_{\gamma\gamma}$ the metric is finite for $|\lambda|>1$, but for $|\lambda|\leq 1$ we have to be careful because the equation
\begin{align*}
\cos(k)=\lambda
\end{align*}
will have exactly one solution in the thermodynamical limit. In that case, let $k_*$ be the solution. If $k$ is an allowed momentum in a small neighborhood of $k_*$, i.e., $k=k_*+\delta q$, then we may write, $\cos(k)\approx \lambda-\sin(k_*)\delta k=\lambda-\sqrt{1-\lambda^2}\delta k$ and $\sin(k)\approx \sin(k_*)+\lambda\delta k=\sqrt{1-\lambda^2} +\lambda \delta k$ (where the sine is positive, since $k\in[0,\pi])$). Then,
\begin{align*}
\frac{\sin^2(k)}{\left(\lambda -\cos(k)\right)^2}\sim \frac{(1-\lambda^2)}{(1-\lambda^2)\delta k^2}\sim N^2,
\end{align*}
so we see that
\begin{align*}
g_{\gamma\gamma}(\lambda,0)\sim N^2, \text{ as } N\to\infty, \text{ for } |\lambda|<1.
\end{align*}
For $|\lambda|=1$ then $\cos(k)=\pm 1 +O(\delta k^2)$ and $\sin(k)\approx -\delta k$ (with $\delta k<0$ since the sine is positive). Hence
\begin{align*}
\frac{\sin^2(k)}{\left(\lambda -\cos(k)\right)^2}\sim \frac{\delta k^2}{\delta k^4}\sim \frac{1}{\delta k^2} \sim N^2,
\end{align*}
as before.

\subsection{Line $\lambda =1$}
For the critical submanifold defined by $\lambda=1$, with tangent vector $\frac{\partial}{\partial \gamma}$, we have

\begin{align*}
g_{\lambda\lambda}(1,\gamma)=\frac{1}{4}\sum_{k}\frac{\gamma^2\sin^2(k)}{\left[\left(1 -\cos(k)\right)^2 +\gamma^2 \sin^2k\right]^2},
\end{align*}
we have to look at momenta in the neighbourhood of $k=0$, i.e., $k=\delta k$ (with $\delta k>0$), and then
\begin{align*}
\sin(k)\approx \delta k \text{ and } \cos(k)\approx 1 -\frac{1}{2}\delta k^2,
\end{align*}
so 
\begin{align*}
\frac{\gamma^2\sin^2(k)}{\left[\left(1 -\cos(k)\right)^2 +\gamma^2 \sin^2k\right]^2}\approx \frac{\gamma^2\delta k^2}{\left[\frac{1}{4}\delta k^4 +\gamma^2 \delta k^2\right]^2}\approx \frac{1}{\gamma^2\delta k^2}\sim \frac{N^2}{\gamma^2},
\end{align*}
thus
\begin{align*}
g_{\lambda\lambda}(1,\gamma)\sim \frac{N^2}{\gamma^2}.
\end{align*}
This means that as $N\to\infty$ it blows up.
Now,
\begin{align*}
g_{\gamma\gamma}(1,\gamma)=\frac{1}{4}\sum_{k}\frac{\left(1-\cos(k)\right)^2\sin^2(k)}{\left[\left(1 -\cos(k)\right)^2 +\gamma^2 \sin^2k\right]^2}.
\end{align*}
Looking at the momenta in the neighborhood of $k=0$ as before, we see that
\begin{align*}
\frac{\left(1-\cos(k)\right)^2\sin^2(k)}{\left[\left(1 -\cos(k)\right)^2 +\gamma^2 \sin^2k\right]^2}\approx \frac{1}{4}\frac{\delta k^6}{\left(\frac{1}{4}\delta k^4 +\gamma^2 \delta k^2\right)^2}\approx \frac{1}{4\gamma^4}\delta k^2,
\end{align*}
so 

\begin{align*}
g_{\gamma\gamma}(1,\gamma) \text{ is  finite} \text{ for } \gamma\neq 0 \text{  and } g_{\gamma\gamma}(1,\gamma)\sim \frac{1}{N^2\gamma^4} \text{ as } \gamma\to 0.
\end{align*}

\subsection{Line $\lambda =-1$}
For the critical submanifold defined by $\lambda=-1$, with tangent vector $\frac{\partial}{\partial \gamma}$, we have

\begin{align*}
g_{\lambda\lambda}(-1,\gamma)=\frac{1}{4}\sum_{k}\frac{\gamma^2\sin^2(k)}{\left[\left(1 +\cos(k)\right)^2 +\gamma^2 \sin^2k\right]^2}.
\end{align*}
We have to look at momenta in the neighborhood of $k=\pi$, i.e., $k=\pi +\delta k$ (with $\delta k<0$), and then
\begin{align*}
\sin(k)\approx -\delta k \text{ and } \cos(k)\approx -1 +\frac{1}{2}\delta k^2,
\end{align*}
so 
\begin{align*}
\frac{\gamma^2\sin^2(k)}{\left[\left(1 +\cos(k)\right)^2 +\gamma^2 \sin^2k\right]^2}\approx \frac{\gamma^2\delta k^2}{\left[\frac{1}{4}\delta k^4 +\gamma^2 \delta k^2\right]^2}\approx \frac{1}{\gamma^2\delta k^2}\sim \frac{N^2}{\gamma^2},
\end{align*}
thus
\begin{align*}
g_{\lambda\lambda}(1,\gamma)\sim \frac{N^2}{\gamma^2}.
\end{align*}
This means that as $N\to\infty$ it blows up.
Now,
\begin{align*}
g_{\gamma\gamma}(-1,\gamma)=\frac{1}{4}\sum_{k}\frac{\left(1+\cos(k)\right)^2\sin^2(k)}{\left[\left(1 +\cos(k)\right)^2 +\gamma^2 \sin^2k\right]^2}.
\end{align*}
Looking at the momenta in the neighborhood of $k=\pi$ as before, we see that
\begin{align*}
\frac{\left(1+\cos(k)\right)^2\sin^2(k)}{\left[\left(1 +\cos(k)\right)^2 +\gamma^2 \sin^2k\right]^2}\approx \frac{1}{4}\frac{\delta k^6}{\left(\frac{1}{4}\delta k^4 +\gamma^2 \delta k^2\right)^2}\approx \frac{1}{4\gamma^4}\delta k^2,
\end{align*}
so 
\begin{align*}
g_{\gamma\gamma}(-1,\gamma) \text{ is  finite} \text{ for } \gamma\neq 0 \text{  and } g_{\gamma\gamma}(-1,\gamma)\sim \frac{1}{N^2\gamma^4} \text{ as } \gamma\to 0.
\end{align*}

\subsection{Summary}
\begin{itemize}
\item [(i)] For critical line $\gamma=0$ and tangent vector $\partial/\partial\lambda$,
\begin{align*}
g_{\lambda\lambda}(\lambda,0)=0 \text{ and } g_{\gamma\gamma}(\lambda,0)=\begin{cases}
\sim N^2 \text{ as } N\to\infty, \text{ for } |\lambda|\leq 1\\
\text{ finite} \text{ otherwise}.
\end{cases}.
\end{align*}
\item [(ii)] For critical line $\lambda=1$ and tangent vector $\partial/\partial \gamma$,
\begin{align*}
g_{\lambda\lambda}(1,\gamma)\sim \frac{N^2}{\gamma^2}\text{ and } g_{\gamma\gamma}(1,\gamma)=\begin{cases} \text{finite} \text{ for } \gamma\neq 0\\ \sim \frac{1}{N^2\gamma^4} \text{ as } \gamma\to 0.\end{cases}.
\end{align*}
\item For critical line $\lambda=-1$ and tangent vector $\partial/\partial \gamma$
\begin{align*}
g_{\lambda\lambda}(-1,\gamma)\sim \frac{N^2}{\gamma^2}\text{ and } g_{\gamma\gamma}(-1,\gamma)=\begin{cases} \text{finite} \text{ for } \gamma\neq 0\\ \sim \frac{1}{N^2\gamma^4} \text{ as } \gamma\to 0.\end{cases}.
\end{align*}
\end{itemize}
In this case we see that $g_{*}=\lim_{N\to\infty} \frac{g}{N}$ restricted to the critical lines vanishes precisely on tangent vectors to the critical lines, except on the intersections $(\lambda,\gamma)=(\pm 1,0)$, where it blows up in all directions. On the complementary subspaces to the tangent spaces it always blows up. In this case there are no operators which are not sufficiently relevant.
\end{widetext}

\end{document}